\documentstyle[epsfig,12pt]{article}
\newcommand{\be}{\begin{eqnarray}}
\newcommand{\ee}{\end{eqnarray}}

\def\lsim{\mathrel{\rlap{\lower4pt\hbox{\hskip1pt$\sim$}}
\raise1pt\hbox{$<$}}}               
\def\gsim{\mathrel{\rlap{\lower4pt\hbox{\hskip1pt$\sim$}}
\raise1pt\hbox{$>$}}}               

\begin{document}

\rightline{\Large{Preprint RM3-TH/00-8}}

\vspace{2cm}

\begin{center}

\Large{SU(6) breaking effects in the nucleon elastic\\ electromagnetic 
form factors}

\vspace{1cm}

\large{Fabio Cardarelli and Silvano Simula}

\vspace{0.5cm}

\normalsize{Istituto Nazionale di Fisica Nucleare, Sezione Roma III\\ Via della Vasca Navale 84, I-00146 Roma, Italy}

\end{center}

\vspace{1cm}

\begin{abstract}

\noindent The effects of both kinematical and dynamical $SU(6)$ breaking on the nucleon elastic form factors, $G_E^N(Q^2)$ and $G_M^N(Q^2)$, are investigated within the constituent quark model formulated on the light-front. The investigation is focused on $G_E^n(Q^2)$ and the ratio $G_M^p(Q^2) / G_M^n(Q^2)$, which within the $SU(6)$ symmetry are given by $G_E^n(Q^2) = 0$ and $G_M^p(Q^2) /$ $G_M^n(Q^2) = - 3/ 2$, respectively. It is shown that the kinematical $SU(6)$ breaking caused by the Melosh rotations of the quark spins as well as the dynamical $SU(6)$ breaking due to the mixed-symmetry component generated in the nucleon wave function by the spin-dependent terms of the quark-quark interaction, can affect both $G_E^n(Q^2)$ and $G_M^p(Q^2) / G_M^n(Q^2)$. The calculated $G_E^n(Q^2)$ is found to be qualitatively consistent with existing data, though only $\simeq 65 \%$ of the experimental neutron charge radius can be explained without invoking effects from possible non-vanishing sizes of the constituent quarks and/or many-body currents. At the same time the predictions for the magnetic ratio $G_M^p(Q^2) / G_M^n(Q^2)$ turn out to be inconsistent with the data. It is however shown that the calculations of $G_M^N(Q^2)$ based on different components of the one-body electromagnetic current lead to quite different results. In particular, the calculations based on the {\em plus} component are found to be plagued by spurious effects related to the loss of the rotational covariance in the light-front formalism. These unwanted effects can be avoided by considering the transverse {\em y}-component of the current. In this way our light-front predictions are found to be consistent with the data on both $G_E^n(Q^2)$ and $G_M^p(Q^2) / G_M^n(Q^2)$. Finally, it is shown that a suppression of the ratio $G_E^p(Q^2) / G_M^p(Q^2)$ with respect to the dipole-fit prediction can be expected in the constituent quark model provided the relativistic effects generated by the Melosh rotations of the constituent spins are taken into account.

\end{abstract}

\newpage

\pagestyle{plain}

\section{Introduction}

\indent The nucleon elastic electromagnetic (e.m.) form factors contain important pieces of information on the internal structure of the nucleon, and therefore an extensive program for their experimental determination is currently undergoing and planned at several facilities around the world \cite{Petratos}. In this work we focus on the nucleon Sachs form factors, defined as \cite{Sachs}
 \be
    G_E^N(Q^2) & = & F_1^N(Q^2) - {Q^2 \over 4M^2} F_2^N(Q^2) ~ ,
    \nonumber \\[3mm]
    G_M^N(Q^2) & = & F_1^N(Q^2) + F_2^N(Q^2)
    \label{eq:sachs}
 \ee
where $F_1^N(Q^2)$ [$F_2^N(Q^2)$] is the Dirac [Pauli] form factor appearing in the covariant decomposition of the (on-shell) nucleon e.m. current matrix elements, viz.
 \be
     <N(p',s')| j_{em}^{\mu}(0) |N(p,s)> = \bar{u}(p',s') \left \{ 
     F_1^N(Q^2) \gamma^{\mu} + F_2^N(Q^2) (i \sigma^{\mu \nu} q_{\nu} / 2M) 
     \right \} u(p,s)
    \label{eq:F1F2}
 \ee
with $Q^2 = - q \cdot q$ and $M$ being the squared four-momentum transfer and the nucleon mass, respectively. As it is well known  (cf., e.g.,  Ref. \cite{SIM99}), the spin-flavour $SU(6)$ symmetry predicts $G_E^n(Q^2) = 0$ and $G_M^p(Q^2) / G_M^n(Q^2) = - 3 / 2$.

\indent The term $(- Q^2 / 4M^2) F_2^N(Q^2)$, appearing in the definition of $G_E^N(Q^2)$ (Eq. (\ref{eq:sachs})), is usually referred to as the Foldy contribution and it is of relativistic origin. Its contribution to the neutron charge radius is known to yield almost totally the experimental value of the latter ($\langle r \rangle_{n, exp}^2 = -0.113 \pm 0.005 ~ fm^2$ from Ref. \cite{radius_n}). This result has been viewed \cite{Weise} as an indication of the smallness of the {\em intrinsic} charge radius related to the neutron rest-frame charge distribution. Very recently \cite{Isgur} however the interpretation of the neutron charge radius as arising from its internal charge distribution has been addressed again, arguing that, going beyond the non-relativistic limit when the Foldy term firstly appears, the Dirac neutron form factor $F_1^n(Q^2)$ receives a relativistic correction that cancels exactly against the Foldy term in $G_E^n(Q^2)$. Such a statement was inferred from the observation that the well-known phenomenon of {\em zitterbewegung}, which produces the Foldy term, cannot contribute to the neutron charge radius, because the latter has zero total charge \cite{Isgur}.

\indent The result of Ref. \cite{Isgur} has been recently derived from a relativistic calculation of the nucleon elastic e.m. form factors, performed within the light-front ($LF$) formalism and the constituent quark ($CQ$) model \cite{SIM99}. Namely it has been shown that the {\em zitterbewegung} approximation of Ref. \cite{Isgur} corresponds to neglect the initial transverse motion in the Melosh rotations of the $CQ$ spins. In Ref. \cite{SIM99} the predictions of both the non-relativistic ($NR$) limit and the {\em zitterbewegung} ($ZB$) approximation have been carried out assuming a pure $SU(6)$-symmetric nucleon wave function. In both cases it was found that $G_E^n(Q^2) = 0$ and $G_M^p(Q^2) / G_M^n(Q^2) = - 3 / 2$. While the latter is consistent with experimental data, the former prediction is well known to be at variance with existing data. Therefore, both in the $NR$ limit and in the $ZB$ approximation the only way to produce a $SU(6)$ breaking can be of dynamical origin, i.e. related to the presence of a mixed-symmetry component generated in the nucleon wave function by the spin-dependent terms of the effective quark-quark potential. We point out that spin-dependent forces are important at short interquark distances, and their strengths are almost fixed by the spin splittings in the hadron mass spectroscopy (cf., e.g., Refs. \cite{CI86} and \cite{Glozmann}). 

\indent However, in Ref. \cite{SIM99} it has been shown that:

\begin{itemize}

\item{the effects of the $CQ$ initial transverse motion in the Melosh rotations can be neglected to a good approximation only when the average value of the transverse momenta, $\langle p_{\perp} \rangle$, is much smaller than the $CQ$ mass $m$. However, in $QCD$ both $m$ and $\langle p_{\perp} \rangle$ are expected to be at least of the order of the $QCD$ scale, $\Lambda_{QCD} \sim 300 ~ MeV$. Moreover, in quark potential models, like those of Refs. \cite{CI86} and \cite{Glozmann}, $\langle p_{\perp} \rangle$ turns out to be much larger than $m$, because of the high momentum components generated in the hadron wave functions by the short-range part of the effective $CQ$ interaction (cf. Ref. \cite{CAR99});}

\item{the Melosh rotations produce a kinematical $SU(6)$ breaking. Indeed, such rotations, being momentum and spin dependent, produce a re-coupling of the $CQ$ orbital angular momentum and spin; in other words, the nucleon $LF$ wave function cannot be any more expressed as a product of a spatial part times a spin-isospin function, and therefore it cannot be $SU(6)$ symmetric. The inclusion of the effects of the $CQ$ initial transverse motion lead unavoidably to $G_E^n(Q^2) \neq 0$ and $G_M^p(Q^2) / G_M^n(Q^2) \neq - 3 / 2$, even when the dynamical $SU(6)$ breaking, generated by the spin-dependent components of the effective $CQ$ interaction, is neglected. Moreover, it has been shown that $LF$ relativistic effects may contribute significantly to $G_E^n(Q^2)$ when $\langle p_{\perp} \rangle > m$, and in particular the calculated neutron charge radius can reach $\sim 40 \%$ of its experimental value;}

\item{the $SU(6)$ breaking associated to the Melosh rotations heavily
affects also the ratio $G_M^p(Q^2) / G_M^n(Q^2)$, leading however to a significative underestimation of the experimental data.}

\end{itemize}

\indent The aim of the present work is twofold: ~ i) to include the effects of the mixed-symmetry wave function both in the $NR$ limit and in the $ZB$ approximation as well as in the full $LF$ approach, extending in this way the analysis of Ref. \cite{SIM99}; ~ ii) to show that $LF$ calculations of $G_M^N(Q^2)$, based on different components of the one-body e.m. current, lead to quite different results, because of spurious, unphysical effects related to the loss of the rotational covariance in the light-front formalism, while the same does not occur in case of the nucleon charge form factor $G_E^N(Q^2)$.

\indent It will be shown that: ~ i) the predictions of the $NR$ limit and the $ZB$ approximation underestimate significantly the data on $G_E^n(Q^2)$, even when the effects of the mixed-symmetry wave function are considered, and ~ ii) once spurious effects are properly avoided, the $LF$ predictions are consistent with the data on both $G_E^n(Q^2)$ and $G_M^p(Q^2) / G_M^n(Q^2)$, though only $\simeq 65 \%$ of the experimental neutron charge radius can be explained without invoking effects from possible non-vanishing $CQ$ sizes and/or many-body currents. This means that the dynamical $SU(6)$ breaking, predicted by quark potential models based on the hadron mass spectroscopy, appears to be consistent with the $SU(6)$ breaking observed in the nucleon elastic e.m. form factors, provided relativistic effects are properly taken into account.

\indent Finally, we will address also the issue of the interpretation of the recently observed \cite{JLAB} deviation of the ratio $\mu_p G_E^p(Q^2) / G_M^p(Q^2)$ from the dipole-fit expectation $\mu_p G_E^p(Q^2) / G_M^p(Q^2) = 1$, where $\mu_p = G_M^p(0)$ is the proton magnetic moment. It will be shown that a suppression of the above mentioned ratio from unity can be expected in the $CQ$ model provided the relativistic effects generated by the Melosh rotations of the $CQ$ spins are taken into account. The results of the calculations obtained using the nucleon wave functions arising from two of the most sophisticated quark potential models \cite{CI86,Glozmann}, are found to compare quite favourably against the recent data \cite{JLAB} from Jefferson Lab ($JLAB$).

\section{The nucleon light-front wave function}

\indent Let us briefly recall the basic notations and the relevant structure
of the nucleon wave function in the $LF$ formalism. As it is well known (cf., e.g., Refs. \cite{KP91,CAR95,CK95}), the nucleon $LF$ wave function is eigenstate of the non-interacting ($LF$) angular momentum operators $j^2$ and $j_n$, where the unit vector $\hat{n} = (0, 0, 1)$ defines the spin quantization axis. The squared free-mass operator is given by
 \be
    M_0^2 = \sum_{i = 1}^3 (|\vec{k}_{i \perp}|^2 + m^2) / \xi_i
    \label{eq:M0_LF}
 \ee
where
 \be
    \xi_i & = & {p_i^+ \over P^+} ~ , \nonumber \\[3mm]
    \vec{k}_{i \perp} & = & \vec{p}_{i \perp} - \xi_i \vec{P}_{\perp}
    \label{eq:LF_var}
 \ee
are the intrinsic $LF$ variables. The subscript $\perp$ indicates the projection perpendicular to the spin quantization axis and the {\em plus} component of a 4-vector $p \equiv (p^0, \vec{p})$ is given by $p^+ = p^0 + \hat{n} \cdot \vec{p}$; finally $\tilde{P} \equiv (P^+, \vec{P}_{\perp}) = \tilde{p}_1 + \tilde{p}_2 + \tilde{p}_3$ is the nucleon $LF$ momentum and $\tilde{p}_i$ the quark one. In terms of the longitudinal momentum $k_{in}$, related to the variable $\xi_i$ by
 \be
    k_{in} \equiv {1 \over 2} \left[ \xi_i M_0 - \left( |\vec{k}_{i 
    \perp}|^2 + m^2 \right) / \xi_i M_0 \right]
    \label{eq:kin}
 \ee
the free mass operator acquires a familiar form, viz. 
 \be
     M_0 = \sum_{i = 1}^3 E_i =  \sum_{i = 1}^3 \sqrt{m^2 + |\vec{k}_i|^2}
    \label{eq:M0}
 \ee
with the three-vectors $\vec{k}_i$ defined as\footnote{Note that $\vec{k}_i$ are internal variables satisfying $\vec{k}_1 + \vec{k}_2 + \vec{k}_3 = 0$.}
 \be
   \vec{k}_i \equiv ( \vec{k}_{i \perp}, k_{in})
   \label{eq:ki}
 \ee
Disregarding the colour variables, the nucleon $LF$ wave function reads as
 \be
    \langle \{ \xi_i \vec{k}_{i \perp}; \nu'_i \tau_i \}|
    \Psi_{N}^{\nu_N}\rangle ~ = ~ \sqrt{{E_1 E_2 E_3 \over M_0 \xi_1 \xi_2 
    \xi_3}} ~ \sum_{ \{\nu_i \}} ~ \langle \{ \nu'_i \} |
    {\cal{R}}^{\dag}|\{\nu_i \} \rangle \langle \{\vec{k_i}; \nu_i \tau_i \} 
    | \chi_{N}^{\nu_N} \rangle 
    \label{eq:wfLF} 
 \ee 
where the curly braces $\{ ~~ \}$ mean a list of indexes corresponding to $i = 1, 2, 3$, and $\nu_i$ ($\tau_i$) is the third component of the $CQ$ spin (isospin). The rotation  ${\cal{R}}^{\dag}$, appearing in Eq. (\ref{eq:wfLF}), is the product of individual (generalized) Melosh  rotations, viz.
 \be
     {\cal{R}}^{\dag} = \prod_{j = 1}^3  R_j^{\dag}(\vec{k}_{j \perp}, 
     \xi_j, m)
     \label{eq:Rmelosh}
 \ee
where
 \be
    R_j(\vec{k}_{j \perp}, \xi_j, m) \equiv {m + \xi_j M_0 - i 
    \vec{\sigma}^{(j)} \cdot (\hat{n} \times \vec{k}_{j \perp}) \over 
    \sqrt{(m + \xi_j M_0)^2 + |\vec{k}_{j \perp}|^2}}
    \label{eq:melosh}
 \ee
with $\vec{\sigma}$ being the ordinary Pauli spin matrices.

\indent In what follows we neglect the very small $P$- and $D$-waves of the nucleon (see later on), and therefore we limit ourselves to the following canonical (or equal-time) wave function (corresponding to a total orbital angular momentum equal to $L =0$):
 \be
    \langle \{ \vec{k}_i; \nu_i \tau_i\}| \chi_{N}^{\nu_N} \rangle & = & 
    w_S(\vec{k}, \vec{p}) ~ {1 \over \sqrt{2}}~ \left[ \Phi^{00}_{\nu_N 
    \tau_N}(\{\nu_i \tau_i \}) + \Phi^{11}_{\nu_N \tau_N}(\{ \nu_i 
    \tau_i \}) \right] + \nonumber \\
    & & w_{S'_s}(\vec{k}, \vec{p}) ~ {1 \over \sqrt{2}} ~ \left[ 
    \Phi^{00}_{\nu_N \tau_N}(\{\nu_i \tau_i \}) - \Phi^{11}_{\nu_N 
    \tau_N}(\{ \nu_i \tau_i \}) \right] + \nonumber \\
    & & w_{S'_a}(\vec{k}, \vec{p}) ~ {1 \over \sqrt{2}} ~ \left[ 
    \Phi^{01}_{\nu_N \tau_N}(\{\nu_i \tau_i \}) + \Phi^{10}_{\nu_N 
    \tau_N}(\{ \nu_i \tau_i \}) \right] + \nonumber \\
    & & w_{A}(\vec{k}, \vec{p}) ~ {1 \over \sqrt{2}} ~ \left[ 
    \Phi^{01}_{\nu_N \tau_N}(\{\nu_i \tau_i \}) - \Phi^{10}_{\nu_N 
    \tau_N}(\{ \nu_i \tau_i \}) \right]
    \label{eq:canonical}
 \ee
where $w_S(\vec{k}, \vec{p})$, $w_{S'_s}(\vec{k}, \vec{p})$, $w_{S'_a}(\vec{k}, \vec{p})$ and $w_A(\vec{k}, \vec{p})$ are the completely symmetric ($S$), the two mixed-symmetry ($S'$) and the completely antisymmetric ($A$) wave functions, respectively. The variables $\vec{k}$ and $\vec{p}$ are the Jacobian internal coordinates, defined as
 \be
   \vec{k} & = & {\vec{k}_1 - \vec{k}_2 \over 2} ~ , 
   \nonumber \\[3mm]
   \vec{p} & = & {2 \vec{k}_3 - \left( \vec{k}_1 + \vec{k}_2 \right) \over 
    3}
   \label{eq:jacobian}
 \ee
with $\vec{k}_i$ given by Eq. (\ref{eq:ki}). Finally, the spin-isospin function $\Phi^{S_{12} T_{12}}_{\nu_N \tau_N}(\{ \nu_i\tau_i \})$, corresponding to a total spin $(1/2)$ and total isospin $(1/2)$, is defined as
 \be
    \Phi^{S_{12} T_{12}}_{\nu_N\tau_N}(\{ \nu_i\tau_i \}) & = & \sum_{M_S} 
    \langle {1 \over 2} \nu_1 {1 \over 2} \nu_2 | S_{12} M_S \rangle ~ 
    \langle S_{12} M_S {1 \over 2} \nu_3 | {1 \over 2} \nu_N \rangle \cdot
    \nonumber \\
    & & \sum_{M_T} \langle {1 \over 2} \tau_1 {1 \over 2} \tau_2 | T_{12} 
    M_T \rangle ~ \langle T_{12} M_T {1 \over 2} \tau_3 | {1 \over 2} 
    \tau_N \rangle 
    \label{eq:STwf}
 \ee
where $S_{12}$ ($T_{12}$) is the total spin (isospin) of the quark pair $(1, 2)$. The normalizations of the various partial waves in Eq. (\ref{eq:canonical}) are
 \be
    \int d\vec{k} ~ d\vec{p} \left| w_S(\vec{k}, \vec{p}) \right|^2 & = & 
    P_S ~ ,
    \nonumber \\[3mm]
    \int d\vec{k} ~ d\vec{p} \left| w_{S'_s}(\vec{k}, \vec{p}) \right|^2 & = 
    & \int d\vec{k} ~ d\vec{p} \left| w_{S'_a}(\vec{k}, \vec{p}) \right|^2 = 
    P_{S'} / 2 ~ , \nonumber \\[3mm]
    \int d\vec{k} ~ d\vec{p} \left| w_A(\vec{k}, \vec{p}) \right|^2 & = & 
    P_A ~ ,
    \label{eq:norms}
 \ee
with $P_S + P_{S'} + P_A = 1$.

\indent Generally speaking, the nucleon $LF$ wave function $|\Psi_{N}^{\nu_N} \rangle$ (see Eq. (\ref{eq:wfLF})) is eigenfunction of a mass equation of the form
 \be
      \left\{ M_0 + V \right\} |\Psi_{N}^{\nu_N} \rangle = M 
      |\Psi_{N}^{\nu_N} \rangle
      \label{eq:mass_LF}
 \ee
where $M_0$ is the free-mass operator (\ref{eq:M0_LF}) and $V$ is a Poincar\'e-invariant interaction term. Using the shorthand notation $|\Psi_{N}^{\nu_N} \rangle = {\cal{R}^{\dag}} |\chi_{N}^{\nu_N} \rangle$, where $|\chi_{N}^{\nu_N} \rangle$ is the nucleon canonical wave function (see Eq. (\ref{eq:canonical})), one gets
 \be
      \left\{ \sum_{i = 1}^3 \sqrt{m^2 + |\vec{k}_i|^2} + {\cal{V}} \right\} 
      |\chi_{N}^{\nu_N} \rangle = M |\chi_{N}^{\nu_N} \rangle
      \label{eq:mass_canonical}
 \ee
where ${\cal{V}} = {\cal{R}} V {\cal{R}^{\dag}}$ is the Melosh-rotated interaction term, while the free-mass $M_0$ is invariant under Melosh rotations and it has been expressed directly in terms of the three-vectors $\vec{k}_i$ through Eq. (\ref{eq:M0}). The Poincar\'e-invariance of $V$ implies simply that ${\cal{V}}$ has to be independent of the total momentum of the system and invariant upon spatial rotations and translations. Therefore, one can adopt for ${\cal{V}}$ any quark potential model able to reproduce the hadron mass spectra, simply interpreting $\vec{k}_i$ as {\em canonical} variables and treating the kinetic energy operator in its relativistic form.

\indent In this work we consider one of the most sophisticated quark potential models, based on the assumption of the valence + gluon dominance (the $OGE$ model of Ref. \cite{CI86}). The baryon (canonical) wave functions are calculated by solving the corresponding three-quark Hamiltonian problem through a variational technique, based on the expansion of the wave function into the harmonic oscillator basis (see appendix A for more details). The coefficients of the expansion and the eigenvalues of the three-quark Hamiltonian are determined via the Raleigh-Ritz variational principle. The presence of spin-orbit and tensor components in the $OGE$ model give rise to non-vanishing $P$- and $D$-waves in the nucleon, respectively. The probabilities of the various partial waves in the nucleon turn out be: $P_S = 98.04 \%$, $P_{S'} = 1.70 \%$, $P_{A} < 10^{-2} \%$, $P_P = 0.05 \%$ and $P_D = 0.21 \%$. It can be seen that due to the smallness of the spin-orbit and tensor terms of the $OGE$ model (which is required by the smallness of the spin-orbit and tensor splittings in the hadron masses) the $P$- and $D$-waves in the nucleon are weakly coupled to the dominant symmetric $S$-wave. Moreover, also the antisymmetric wave $w_A(\vec{k}, \vec{p})$ exhibits an extremely weak coupling  to the dominant symmetric $S$-wave. Finally, the probability of the mixed-symmetry $S'$ component is of the order of few $\%$, being governed by the strength of the spin-spin component of the $OGE$ model, which is almost fixed by the $N - \Delta(1232)$ mass splitting. We point out that the same basic features are shared also by the nucleon wave function arising from a recently proposed quark potential model, based on the exchange of the pseudoscalar mesons arising from the spontaneous breaking of chiral symmetry (the chiral model of Ref. \cite{Glozmann}). As a matter of fact, in case of the chiral model the partial wave probabilities are found to be: $P_S = 98.73 \%$, $P_{S'} = 1.27 \%$, $P_{A} < 10^{-2} \%$\footnote{In case of the version of Ref. \cite{Glozmann} the chiral quark potential model does not contain any spin-orbit and tensor terms and therefore the resulting nucleon wave function has no $P$- and $D$-waves.}.

\begin{figure}[htb]

\vspace{0.15cm}

\centerline{\epsfxsize=16cm \epsfig{file=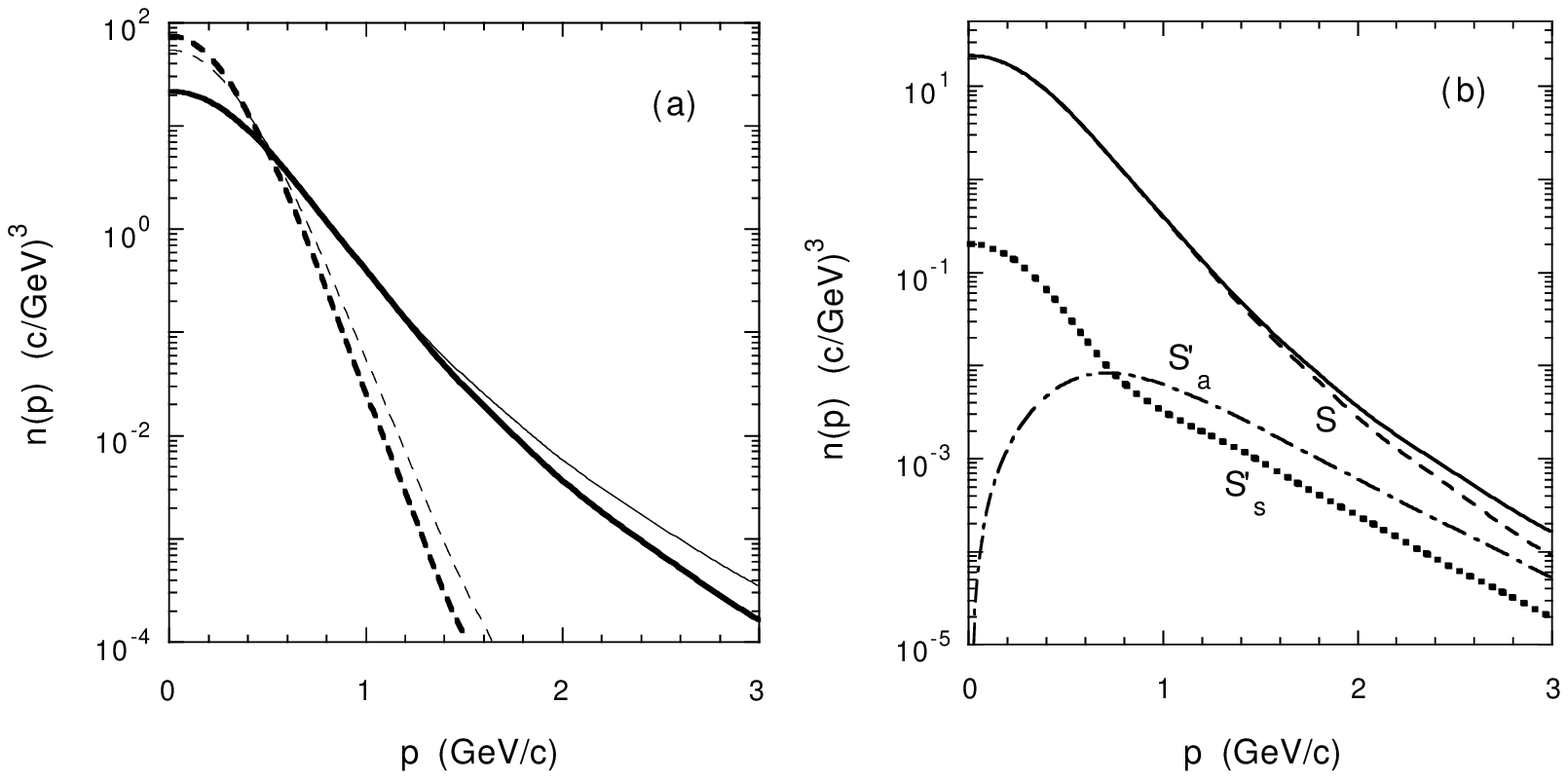}}

{\small {\bf Figure 1.} Canonical $CQ$ momentum distribution in the nucleon, $n(p)$ (Eq. (\protect\ref{eq:n(p)})), versus the three-momentum $p = |\vec{p}|$ (see Eq. (\protect\ref{eq:jacobian})).  (a) Thick and thin lines correspond to the nucleon eigenfunctions of the $OGE$ \protect\cite{CI86} and chiral \cite{Glozmann} quark potential models, respectively. Solid lines are the results obtained using the full interaction models, while dotted lines correspond to the inclusion of their linear confinement terms only. (b) Contributions of various partial waves (see Eq. (\protect\ref{eq:canonical})) to the $CQ$ momentum distribution $n(p)$ obtained within the $OGE$ interaction model. Dot-dashed, dotted and dashed lines correspond to $S'_a$-, $S'_s$- and $S$-waves, respectively. The solid line is their sum.}

\vspace{0.15cm}

\end{figure}

\indent In Ref. \cite{CAR99} the wave functions of the nucleon and of the most prominent electroproduced nucleon resonances corresponding to the $OGE$ and chiral quark potential models have been compared. Here, we have reported in Fig. 1(a) the (canonical) $CQ$ momentum distribution $n(p)$ in the nucleon, calculated within the $OGE$ and chiral models neglecting the small $P$-, $D$- and antisymmetric $w_A $ partial waves, namely:
 \be
       n(p = |\vec{p}|) \equiv \int d\Omega_{\vec{p}} ~ d\vec{k} \left\{ 
       \left|  w_S(\vec{k}, \vec{p}) \right|^2 + \left| w_{S'_s}(\vec{k}, 
       \vec{p}) \right|^2 + \left| w_{S'_a}(\vec{k}, \vec{p}) \right|^2 
       \right\} ~ .
       \label{eq:n(p)}
 \ee

\indent From Fig. 1(a) it can clearly be seen that both quark potential models predict a remarkable content of high-momentum components, which are generated by the short-range part of the quark-quark interaction (cf. also Ref. \cite{CAR99}). In case of the $OGE$ model the interaction at short-range is mainly due to the flavour-independent central Coulomb-like and spin-spin terms arising from the effective one-gluon-exchange, while in case of the chiral model the  interaction at short interquark distances is governed by the spin-flavour structure characterizing the exchanges of pions, $\eta$ and partially $\eta'$ mesons. As a matter of fact, the above mentioned high-momentum components are completely absent if one switches off the intermediate boson exchanges and only the linear confining term (dominant at large interquark distances) is considered. In Fig. 1(b) we have reported the individual contributions of the $S$-, $S'_s$- and $S'_a$- waves to the $CQ$ momentum distribution $n(p)$, calculated within the $OGE$ model. It can be seen that both the $S'_s$ and $S'_a$ components are negligible at low momenta, while they may be important at high momenta, as expected from the short-range nature of the spin-spin interaction among quarks.

\begin{table}[htb]

\vspace{0.15cm}

{\small {\bf Table 1}. Values of the average $CQ$ transverse momentum $\langle p_{\perp} \rangle$ (\ref{eq:p_perp}), the $CQ$ mass radii $\langle r \rangle_{u(d)}$ (\ref{eq:radii}) and the charge radii of proton and neutron (\ref{eq:rpn}) predicted by the $OGE$ \cite{CI86} and the chiral \cite{Glozmann} quark potential models.}

\begin{center}

\begin{tabular}{||c||c|c|c||c|c||}
\hline \hline
 & \multicolumn{3}{c||}{$OGE ~ model$} & \multicolumn{2}{c||}{$chiral ~ model$} \\ \cline{2-6}
 & $conf. ~ only$ & $w/o ~ S'$ & $with ~ S'$  & $w/o ~ S'$ & $with ~ S'$ \\ \hline \hline
 $\langle p_{\perp} \rangle ~ (GeV)$ & $0.33$  & $0.56$ & $0.58$  & $0.60$ & $0.61$ \\ \hline \hline
 $\langle r \rangle_u ~ (fm)$ & $0.49$ & $0.31$ & $0.32$ & $0.31$ & $0.32$ \\ \hline
 $\langle r \rangle_d ~ (fm)$ & $0.49$ & $0.31$ & $0.31$ & $0.31$ & $0.30$ \\ \hline \hline
 $\langle r \rangle_p ~ (fm)$ & $0.49$ & $0.31$ & $0.32$ & $0.31$ & $0.32$ \\ \hline
 $\langle r \rangle_n^2 ~ (fm^2)$  & $0.00$ & $0.00$ & $-0.0083$ & $0.00$ & $-0.0090$ \\ \hline \hline
\end{tabular}

\end{center}

\vspace{0.15cm}

\end{table}

\indent Finally, we have collected in Table 1 the predictions for few observables which are of interest in this work, namely the (canonical) average $CQ$ transverse momentum $\langle p_{\perp} \rangle$
 \be
       \langle p_{\perp} \rangle \equiv \sqrt{{2 \over 3} \int_0^{\infty} dp 
       ~ p^4 ~ n(p)}
       \label{eq:p_perp}
 \ee
and the (canonical) $CQ$ mass radii $\langle r \rangle_{u(d)}$ in the proton
 \be
       \langle r \rangle_{u(d)} \equiv  \sqrt{{1 \over 2} \sum_{\nu_p} 
       \langle \chi_{p}^{\nu_p}| \sum_{i=1}^3 {1 \over 2} \left(1 \pm 
       \tau_3^{(i)} \right) |\vec{r}_i|^2 |\chi_{p}^{\nu_p} \rangle}
       \label{eq:radii}
 \ee
where $\vec{r}_i$ is the conjugate variable to $\vec{k}_i$. In terms of $\langle r \rangle_{u(d)}$ the squared (canonical) charge radii of proton and neutron are given by
 \be
       \langle r \rangle_p^2 & = & {1 \over 3} \left( 4 \langle r 
       \rangle_u^2 - \langle r \rangle_d^2 \right) \nonumber \\[3mm]
       \langle r \rangle_n^2 & = & {2 \over 3} \left( \langle r \rangle_d^2 
       - \langle r \rangle_u^2 \right)
       \label{eq:rpn}
 \ee
From Table 1 it can be seen that: ~ i) both models give rise to a proton with a small size ($\sim 0.3 ~ fm$ instead of $\sim 0.8 ~ fm$), and ~ ii) the predicted neutron charge radius has the correct sign, but its absolute value is about one order of magnitude less than the experimental value, $\langle r \rangle_{n, exp}^2 = -0.113 \pm 0.005 ~ fm^2$ \cite{radius_n}.

\section{Calculations of the nucleon elastic form factors}

\indent As in Ref. \cite{SIM99} (cf. also Ref. \cite{CAR95}) we consider the one-body component of the e.m. current operator at the $CQ$ level, viz.
 \be 
    {\cal{I}}^{\nu} = \sum_{j = 1}^3 I_j^{\nu} = \sum_{j=1}^3 \left[ e_j
    \gamma^{\nu} f_1^j(Q^2) ~ + ~ i \kappa_j {\sigma^{\nu \mu} q_{\mu}
    \over 2m} f_2^j(Q^2) \right]
    \label{eq:current}
 \ee
where $\sigma^{\nu \mu} = {i \over 2}[\gamma^{\nu},\gamma^{\mu}]$, $e_j$ is
the charge of the j-th quark, $\kappa_j$ the corresponding anomalous magnetic moment and $f_{1(2)}^j(Q^2)$ its Dirac (Pauli) form factor (with
$f_{1(2)}^j(Q^2 = 0) = 1$). In the $LF$ formalism the form factors for a conserved current can be obtained using only the matrix elements of the {\em plus} component of the current operator and, moreover, for $Q^2 \geq 0$ the choice of a frame where $q^+ = 0$ allows to suppress the contribution of the $Z$-graph (i.e., pair creation from the vacuum \cite{ZGRAPH}). In what follows we adopt a Breit frame where the four-vector $q \equiv (q^0, \vec{q})$ is given by $q^0 = 0$ and $\vec{q} = (q_x, q_y, q_z) = (Q, 0, 0)$, while the unit vector $\hat{n} = (0, 0, 1)$ defines the spin-quantization axis (as in Section 2). In case of the nucleon one has
 \be 
    \langle \Psi_{N}^{\nu'_N}| ~ {\cal{I}}^+ ~ |\Psi_N^{\nu_N} \rangle =
    F_1^N(Q^2) ~ \delta_{{\nu'}_N \nu_{N}} - i {Q \over 2 M} 
    F_2^N(Q^2) ~ \langle \nu'_{N} | \sigma_y | \nu_{N} \rangle
    \label{eq:Iplus}
 \ee
leading to
 \be
    F_1^N(Q^2) & = & {1 \over 2} \mbox{Tr}\{ {\cal{I}}^+ \} ~ ,
    \nonumber \\[3mm]
    F_2^N(Q^2) & = & {2M \over Q} {1 \over 2} \mbox{Tr}\{ {\cal{I}}^+ 
    i\sigma_y \}
    \label{eq:F1F2+}
 \ee

\indent Using Eqs. (\ref{eq:wfLF}-\ref{eq:STwf}) for the nucleon wave function, the general structure of the Dirac and Pauli nucleon form factors (\ref{eq:F1F2+}), derived from the matrix elements of the {\em plus} component of the one-body e.m. current (\ref{eq:current}), is given explicitly by:
 \be
     F_{\alpha}^{p(n)}(Q^2) & = & 3 \left[ \delta_{\alpha,1} - {2M \over Q} 
    \delta_{\alpha, 2} \right] \int [d\xi] [d\vec{k}_{\perp}] 
    [d\vec{k'}_{\perp}] ~ \sqrt{{E_1 E_2 E_3 {M'}_0 \over {E'}_1 {E'}_2 
    {E'}_3 M_0}}  \cdot \nonumber \\[3mm]
    & &  \sum_{\beta = 1, 2} \left\{ f_{0, ~ \beta}^{p(n)}(Q^2) \left[ 
    {\cal{R}}_{00}^{(\alpha \beta)} ~ \bar{w}_{00}^*(\vec{k'}, \vec{p'}) 
    \bar{w}_{00}(\vec{k}, \vec{p}) + {\cal{R}}_{11}^{(\alpha \beta)} ~ 
    \bar{w}_{10}^*(\vec{k'}, \vec{p'})  \bar{w}_{10}(\vec{k}, \vec{p}) + 
    \right. \right. \nonumber \\[3mm]
    & & \left. \left. {\cal{R}}_{01}^{(\alpha \beta)} ~ 
    \bar{w}_{00}^*(\vec{k'}, \vec{p'}) \bar{w}_{10}(\vec{k}, \vec{p}) + 
    {\cal{R}}_{10}^{(\alpha \beta)} ~ \bar{w}_{10}^*(\vec{k'}, \vec{p'}) 
    \bar{w}_{00}(\vec{k}, \vec{p}) \right] + \right. \nonumber \\[3mm]
    & & \left. f_{1, ~ \beta}^{p(n)}(Q^2) \left[ {\cal{R}}_{11}^{(\alpha 
    \beta)} ~ \bar{w}_{11}^*(\vec{k'}, \vec{p'}) \bar{w}_{11}(\vec{k}, 
    \vec{p}) + {\cal{R}}_{00}^{(\alpha \beta)} ~ \bar{w}_{01}^*(\vec{k'}, 
    \vec{p'})  \bar{w}_{01}(\vec{k}, \vec{p}) + \right. \right. 
    \nonumber \\[3mm]
    & & \left. \left. {\cal{R}}_{10}^{(\alpha \beta)} ~ 
    \bar{w}_{11}^*(\vec{k'}, \vec{p'}) \bar{w}_{01}(\vec{k}, \vec{p}) + 
    {\cal{R}}_{01}^{(\alpha \beta)} ~ \bar{w}_{01}^*(\vec{k'}, \vec{p'}) 
    \bar{w}_{11}(\vec{k}, \vec{p}) \right]  \right\}
     \label{eq:F1F2_LF}
 \ee
where $\alpha, ~ \beta = 1, 2$ and
 \be
   \bar{w}_{00}(\vec{k}, \vec{p}) & = & {1 \over \sqrt{2}} \left[ 
   w_S(\vec{k}, \vec{p}) + w_{S'_s}(\vec{k}, \vec{p}) \right] ~ , 
   \nonumber \\
   \bar{w}_{01}(\vec{k}, \vec{p}) & = & {1 \over \sqrt{2}} \left[ 
   w_{S'_a}(\vec{k}, \vec{p}) + w_A(\vec{k}, \vec{p}) \right] ~ ,
   \nonumber \\
   \bar{w}_{10}(\vec{k}, \vec{p}) & = & {1 \over \sqrt{2}} \left[ 
   w_{S'_a}(\vec{k}, \vec{p}) - w_A(\vec{k}, \vec{p}) \right] ~ ,
   \nonumber \\
   \bar{w}_{11}(\vec{k}, \vec{p}) & = & {1 \over \sqrt{2}} \left[ 
   w_S(\vec{k}, \vec{p}) - w_{S'_s}(\vec{k}, \vec{p}) \right] ~ ,
   \label{eq:wST}
 \ee
with
 \be
   \left[ d\xi \right] & = & d\xi_1 ~ d\xi_2 ~ d\xi_3 ~ \delta[\xi_1 + 
   \xi_2 + \xi_3 - 1] ~ ,
   \nonumber \\[3mm]
   \left[ d\vec{k}_{\perp} \right] & = & d\vec{k}_{1 \perp} ~ 
   d\vec{k}_{2 \perp} ~ d\vec{k}_{3 \perp} ~ \delta[\vec{k}_{1 \perp} + 
   \vec{k}_{2 \perp} + \vec{k}_{3 \perp}] ~ ,
   \nonumber \\[3mm]
   \left[ d\vec{k'}_{\perp} \right] & = & d\vec{k'}_{1 \perp} ~ 
   d\vec{k'}_{2 \perp} ~ d\vec{k'}_{3 \perp} ~ \delta[\vec{k'}_{1 \perp} - 
   \vec{k}_{1 \perp} + \xi_1 \vec{q}_{\perp}] ~ \delta[\vec{k'}_{2 \perp} - 
   \vec{k}_{2 \perp} + \xi_2 \vec{q}_{\perp}] 
   \nonumber \\[3mm]
   & & \delta[\vec{k'}_{3 \perp} - \vec{k}_{3 \perp} + (\xi_3 - 1 ) ~ 
   \vec{q}_{\perp}]
   \label{eq:int}
 \ee
The coefficients ${\cal{R}}_{S'_{12} S_{12}}^{(\alpha \beta)}$, appearing in Eq. (\ref{eq:F1F2_LF}) and containing the effects of the Melosh rotations, are explicitly given in the Appendix B, while the quantities $f_{0, ~ \alpha}^{p(n)}(Q^2)$ and $f_{1, ~ \alpha}^{p(n)}(Q^2)$ are appropriate isospin combinations of the $CQ$ form factors, viz.
 \be
   f_{0, ~ \alpha=1}^{p(n)}(Q^2) & = & e_{u(d)} ~ f_1^{u(d)}(Q^2) ~ ,
   \nonumber \\[3mm]
   f_{0, ~ \alpha=2}^{p(n)}(Q^2) & = & {Q \over 2m} \kappa_{u(d)} ~ 
   f_2^{u(d)}(Q^2) ~ ,\nonumber \\[3mm]
   f_{1, ~ \alpha=1}^{p(n)}(Q^2) & = & {1 \over 3} \left[ e_{u(d)} ~ 
   f_1^{u(d)}(Q^2) + 2 e_{d(u)} ~ f_1^{d(u)}(Q^2) \right] ~ ,
   \nonumber \\[3mm]
   f_{1, ~ \alpha=2}^{p(n)}(Q^2) & = & {1 \over 3} {Q \over 2m} \left[ 
   \kappa_{u(d)} ~ f_2^{u(d)}(Q^2) + 2 \kappa_{d(u)} ~ f_2^{d(u)}(Q^2) 
   \right]
   \label{eq:fST}
 \ee
In case of a $SU(6)$-symmetric nucleon wave function (i.e., $w_{S'_s} = w_{S'_a} = w_A = 0$) it is straightforward to check that Eq. (\ref{eq:F1F2_LF}) reduces to Eqs. (10-11) of Ref. \cite{SIM99}.

\indent We show now the structure of the nucleon Sachs form factors in the $NR$ limit and in the $ZB$ approximation of Ref. \cite{Isgur}. For sake of simplicity we consider the case $w_A = 0$, because, as already observed, the completely antisymmetric partial wave is very weakly coupled with the dominant symmetric $S$-wave. Moreover, as in Ref. \cite{SIM99} we will limit ourselves to the case of flavour symmetric $CQ$ form factors, i.e. 
 \be
   f_1^j(Q^2) & = & f(Q^2) ~ , \nonumber \\[3mm]
   f_2^j(Q^2) & = & \tilde{f}(Q^2) ~ , \nonumber \\[3mm]
   k_j & = & e_j \kappa
   \label{eq:SU2}
 \ee
with $f(Q^2 = 0) = \tilde{f}(Q^2 = 0) = 1$. In the $NR$ limit the Fourier transforms of the non-relativistic charge and magnetization densities yield (cf. Ref. \cite{SIM99}) 
 \be
   G_E^p(Q^2) & = & F_s^{NR}(Q^2) + \Delta_s^{NR}(Q^2) + F_a^{NR}(Q^2) ~ ,
   \nonumber \\[3mm]
   G_M^p(Q^2) & = & 3 \left\{ F_s^{NR}(Q^2) + \Delta_s^{NR}(Q^2) + \kappa 
   \left[ \tilde{F}_s^{NR}(Q^2) + \tilde{\Delta}_s^{NR}(Q^2) \right] 
   \right\} - \nonumber \\[3mm] 
   & & \left[ F_a^{NR}(Q^2) + \kappa \tilde{F}_a^{NR}(Q^2) \right]
   ~ , \nonumber \\[3mm]
   G_E^n(Q^2) & = & - \Delta_s^{NR}(Q^2) ~ ,
   \nonumber \\[3mm]
   G_M^n(Q^2) & = & - 2 \left\{ F_s^{NR}(Q^2) + {1 \over 2} 
   \Delta_s^{NR}(Q^2) + \kappa \left[ \tilde{F}_s^{NR}(Q^2) + {1 \over 2} 
   \tilde{\Delta}_s^{NR}(Q^2) \right] \right\} + \nonumber \\[3mm]
   & & 2 \left[ F_a^{NR}(Q^2) +\kappa \tilde{F}_a^{NR}(Q^2) \right] ~ ,
   \label{eq:GEGM_nr}
 \ee
where binding effects in the nucleon mass have been neglected (i.e. $M = 3m$) and
 \be
   F_s^{NR}(Q^2 = |\vec{q}|^2) & = & f(Q^2) ~ \int d\vec{k} d\vec{p} ~ 
   \left[ w_S^*(\vec{k}, \vec{p} + {2 \over 3} \vec{q}) ~ w_S(\vec{k}, 
   \vec{p}) + (S \rightarrow S'_s) \right] ~ , \nonumber \\
   \Delta_s^{NR}(Q^2 = |\vec{q}|^2) & = & f(Q^2) ~ \int d\vec{k} d\vec{p} ~ 
   \left[ w_S^*(\vec{k}, \vec{p} + {2 \over 3} \vec{q}) ~ w_{S'_s}(\vec{k}, 
   \vec{p}) + (S \leftrightarrow S'_s) \right] ~ , \nonumber \\
    F_a^{NR}(Q^2 = |\vec{q}|^2) & = & f(Q^2) ~ \int d\vec{k} d\vec{p} ~ 
   w_{S'_a}^*(\vec{k}, \vec{p} + {2 \over 3} \vec{q}) ~ w_{S'_a}(\vec{k}, 
   \vec{p}) 
   \label{eq:FD_nr}
 \ee
while $\tilde{F}_s^{NR}(Q^2)$, $\tilde{\Delta}_s^{NR}(Q^2)$ and $\tilde{F}_a^{NR}(Q^2)$ are given by Eq. (\ref{eq:FD_nr}) with $f(Q^2)$ replaced by $\tilde{f}(Q^2)$.
 
\indent The $ZB$ approximation can be worked out from Eq. (\ref{eq:F1F2_LF}) following the procedure explained in Ref. \cite{SIM99} (cf. also Appendix B), which includes again the neglect of binding effects in the nucleon mass (i.e. $M = 3m$) for compatibility with the $NR$ reduction. One obtains
 \be
    G_E^p(Q^2) & = & (1 - {Q^2 \over 2M^2}) \left[ F_s^{ZB}(Q^2) + 
    \Delta_s^{ZB}(Q^2) + F_a^{ZB}(Q^2) \right] ~ , \nonumber \\[3mm]
    G_M^p(Q^2) & = & 3 \left\{ F_s^{ZB}(Q^2) + \Delta_s^{ZB}(Q^2) + \kappa 
    (1 +{Q^2 \over 4M^2}) \left[ \tilde{F}_s^{ZB}(Q^2) + 
    \tilde{\Delta}_s^{ZB}(Q^2) \right] \right\} - \nonumber \\[3mm] 
    & & \left[ F_a^{ZB}(Q^2) + \kappa (1 +{Q^2 \over 4M^2}) 
    \tilde{F}_a^{ZB}(Q^2) \right] ~ , \nonumber \\[3mm]
    G_E^n(Q^2) & = & - (1 - {Q^2 \over 2M^2}) \Delta_s^{ZB}(Q^2) ~ ,
    \nonumber \\[3mm]
    G_M^n(Q^2) & = & -2 \left\{ F_s^{ZB}(Q^2) + {1 \over 2} 
    \Delta_s^{ZB}(Q^2) + \kappa (1 +{Q^2 \over 4M^2}) \left[ 
    \tilde{F}_s^{ZB}(Q^2) + {1 \over 2} \tilde{\Delta}_s^{ZB}(Q^2) \right] 
    \right\} + \nonumber \\[3mm] 
    & & 2 \left[ F_a^{ZB}(Q^2) + \kappa (1 +{Q^2 \over 4M^2}) 
    \tilde{F}_a^{ZB}(Q^2) \right] ~ ,
    \label{eq:GEGM_zb}
 \ee
where
 \be
    F_s^{ZB}(Q^2) & = & {f(Q^2) \over \sqrt{1 + Q^2 / M^2}} \int [d\xi] 
    [d\vec{k}_{\perp}] [d\vec{k'}_{\perp}] ~ J ~ \left[ w_S^*(\vec{k'},  
    \vec{p'}) ~ w_S(\vec{k}, \vec{p}) + (S \rightarrow S'_s) \right] ~ ,
    \nonumber \\
    \Delta_s^{ZB}(Q^2) & = & { f(Q^2) \over \sqrt{1 + Q^2 / M^2}} \int 
    [d\xi] [d\vec{k}_{\perp}]  [d\vec{k'}_{\perp}] ~ J ~ \left[ 
    w_S^*(\vec{k'}, \vec{p'}) ~ w_{S'_s}(\vec{k}, \vec{p}) + (S 
    \leftrightarrow S'_s) \right]  ~ , \nonumber \\
    F_a^{ZB}(Q^2) & = & {f(Q^2) \over \sqrt{1 + Q^2 / M^2}} \int [d\xi] 
    [d\vec{k}_{\perp}] [d\vec{k'}_{\perp}] ~ J ~ w_{S'_a}^*(\vec{k'},  
   \vec{p'}) ~ w_{S'_a}(\vec{k}, \vec{p})
    \label{eq:FD_zb}
 \ee
with $J \equiv \sqrt{E_1 E_2 E_3 {M'}_0 / {E'}_1 {E'}_2 {E'}_3 M_0}$. The quantities $\tilde{F}_s^{ZB}(Q^2)$, $\tilde{\Delta}_s^{ZB}(Q^2)$ and $\tilde{F}_a^{ZB}(Q^2)$ are given by Eq. (\ref{eq:FD_zb}) with $f(Q^2)$ replaced by $\tilde{f}(Q^2)$.

\indent From Eqs. (\ref{eq:GEGM_nr}-\ref{eq:FD_zb}) it can be seen that both in the $NR$ limit and in the $ZB$ approximation $G_E^n(Q^2) \neq 0$ and $G_M^p(Q^2) / G_M^n(Q^2) \neq 3 / 2$ can obtained only when the mixed-symmetry $S'$-wave is considered, more precisely through the interference between the dominant symmetric $S$-wave and the mixed-symmetry $S'_s$-wave (cf. the definitions of $\Delta_s^{NR}(Q^2)$ and $\Delta_s^{ZB}(Q^2)$). Such interference is expected to take place only at high values of the $CQ$ momentum (see Fig. 1(b)). Finally, we stress again that in the full $LF$ calculations (\ref{eq:F1F2_LF}) both $G_E^n(Q^2) \neq 0$ and $G_M^p(Q^2) / G_M^n(Q^2) \neq 3 / 2$ can be obtained simply via relativistic effects only, i.e. without the inclusion of the effects of the mixed-symmetry $S'$-wave.

\section{Results}

\indent We have calculated the nucleon form factors adopting the nucleon wave function of the $OGE$ quark potential model of Ref. \cite{CI86} both including and excluding the effects due to the mixed-symmetry $S'$-wave\footnote{We have obtained results very similar to those presented in this Section adopting also the nucleon wave function of the chiral quark potential model of Ref. \cite{Glozmann}, where the $CQ$ mass has been chosen at the value $m = 0.340 ~ GeV$.}. In the calculations we have employed only the matrix elements of the {\em plus} component of the e.m. current (see Eq. (\ref{eq:F1F2+})), assuming also the case of point-like $CQ$'s in Eq. (\ref{eq:current}) [i.e. putting $f(Q^2) = \tilde{f}(Q^2) = 1$ and $\kappa = 0$ in the right-hand side of Eq. (\protect\ref{eq:SU2})]. The $CQ$ mass $m$ has been always taken at the value $m = 0.220 ~ GeV$ from the $OGE$ model. Moreover, as already pointed out in Section 3, both in the $NR$ limit and in the $ZB$ approximation the binding effects in the nucleon mass have been neglected (i.e. $M = 3m$), whereas in case of the full $LF$ calculations the nucleon mass is taken at its physical value ($M = 0.938 ~ GeV$).

\begin{figure}[htb]

\vspace{0.15cm}

\centerline{\epsfxsize=16cm \epsfig{file=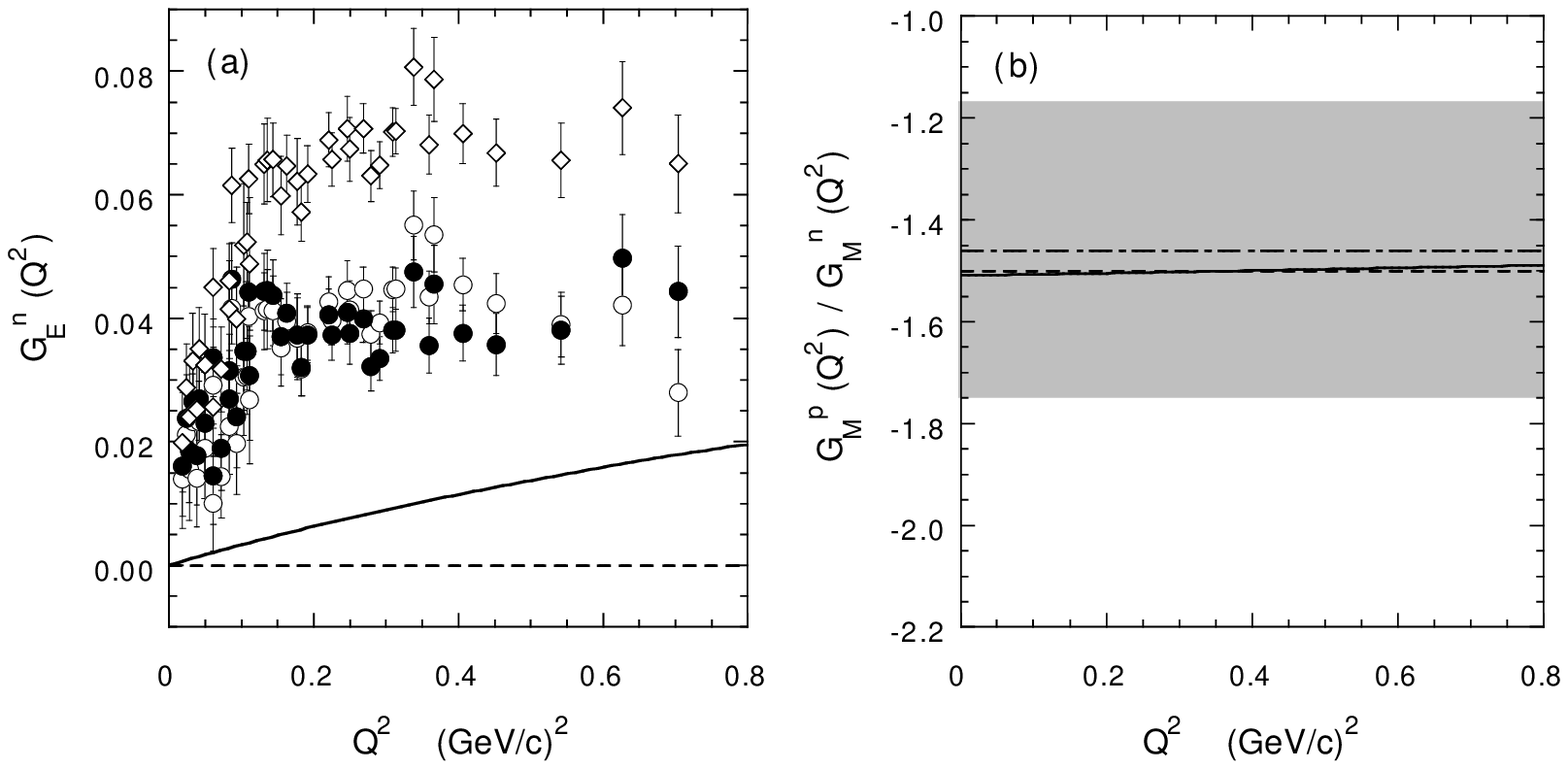}}

{\small {\bf Figure 2.} (a) The neutron charge form factor $G_E^n(Q^2)$ vs. $Q^2$ within the $NR$ limit (see Eq. (\protect\ref{eq:GEGM_nr})), assuming point-like $CQ$'s. The data are from Ref. \cite{Platchkov}, where the Reid Soft Core (open dots), Paris (full dots) and Nijemegen (diamonds) nucleon-nucleon interaction were adopted. (b) The ratio $G_M^p(Q^2) /$ $G_M^n(Q^2)$ vs. $Q^2$. The shaded area corresponds to a $\pm 15 \%$ deviation \cite{SIM99} from the dipole-fit expectation $G_M^p(Q^2) / G_M^n(Q^2) \simeq \mu_p / \mu_n \simeq -1.46$ (dot-dashed line). Solid and dashed lines are the results obtained with and w/o the mixed-symmetry $S'$-wave, respectively.}

\vspace{0.15cm}

\end{figure}

\indent Our results obtained for $G_E^n(Q^2)$ and the ratio $G_M^p(Q^2) / G_M^n(Q^2)$ in case of the $NR$ limit (\ref{eq:GEGM_nr}), the $ZB$ approximation (\ref{eq:GEGM_zb}) and the full $LF$ calculations (\ref{eq:F1F2_LF}), are reported in Figs. 2-4, respectively, and compared with the experimental data. In this respect, let us remind that existing data on $G_M^p(Q^2)$ and $G_M^n(Q^2)$ exhibit a well-known dipole behaviour, leading to $G_M^p(Q^2) / G_M^n(Q^2) \simeq \mu_p / \mu_n \simeq -1.46$ with only a $10 \div 15 \%$ uncertainty up to $Q^2 \sim 1 ~ (GeV/c)^2$ (cf., e.g., Ref. \cite{Petratos}). 

\begin{figure}[htb]

\vspace{0.15cm}

\centerline{\epsfxsize=16cm \epsfig{file=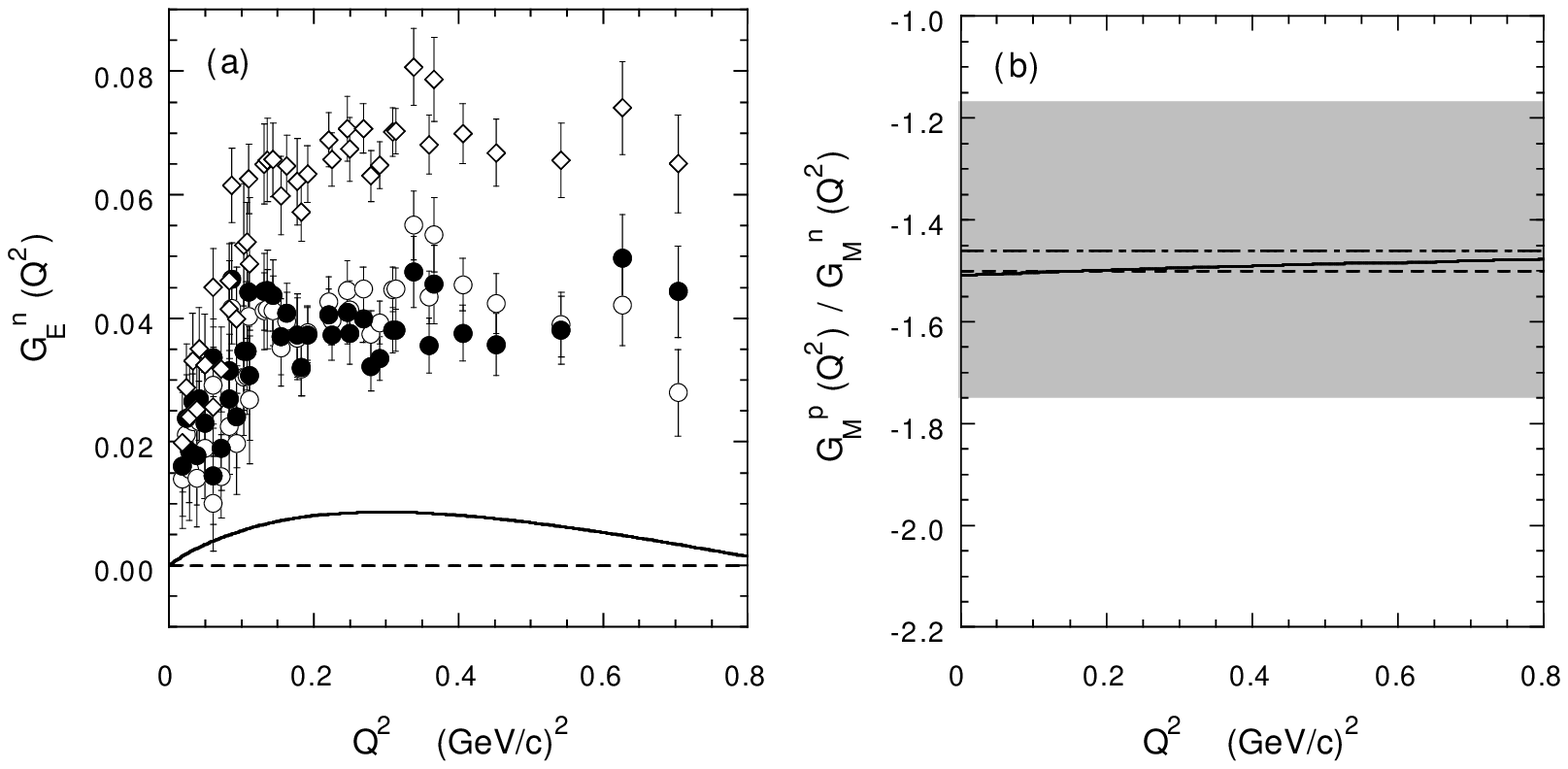}}

{\small \centerline{{\bf Figure 3.} The same as in Fig. 2, but within the $ZB$ approximation (see Eq. (\protect\ref{eq:GEGM_zb})).}}

\vspace{0.15cm}

\end{figure}

\indent From Figs. 2 and 3 it can be seen that both in the $NR$ limit and in the $ZB$ approximation the impact of the $S'$-wave is quite limited. As for $G_E^n(Q^2)$ (see Figs. 2(a) and 3(a)), both the $NR$ and the $ZB$ predictions are quite far from the experimental data. In particular the calculated neutron charge radius is still given by its canonical expectation $\langle r \rangle_n^2 \simeq - 0.01 ~ fm^2$ given in Table 1, which is about one order of magnitude less than the experimental value,  $\langle r \rangle_{n, exp}^2 = -0.113 \pm 0.005 ~ fm^2$ \cite{radius_n}. Such an underestimate is mainly due to the smallness of the $CQ$ mass radii (\ref{eq:radii}) predicted by the $OGE$ (as well as by the chiral) model (cf. Table 1).  Both in the $NR$ limit and in the $ZB$ approximation the calculated $G_E^n(Q^2)$ receives a positive contribution from the inclusion of the $S'$-wave. As pointed out in Ref. \cite{Isgur99}, the sign of this contribution can be expected from the changes that the $S'$-wave produces on the values of $CQ$ mass radii (\ref{eq:radii}). As a matter of fact, the spin-spin force makes the $d$-quark closer to the center-of-mass of the proton than the $u$-quark, so that $\langle r \rangle_d < \langle r \rangle_u$ (cf. Table 1). Thus the neutron charge radius receives a negative contribution (see Eq. (\ref{eq:rpn})), which implies an effect on $G_E^n(Q^2)$ with a positive sign. As for $G_M^p(Q^2) / G_M^n(Q^2)$ (see Figs. 2(b) and 3(b)), both the $NR$ and the $ZB$ predictions are only slightly modified by the inclusion of the mixed-symmetry $S'$-wave and therefore they are still consistent with the experimental data, as already noted in Ref. \cite{SIM99}.

\indent In case of the full $LF$ calculations the effects of the $S'$-wave on $G_E^n(Q^2)$ are more substantial (see Fig. 4(a)). Indeed, the $LF$ prediction receives an important contribution from the inclusion of the $S'$-wave. The sign of this contribution is again positive as in the cases of the $NR$ limit and the $ZB$ approximation, but with a significantly larger impact (cf. Figs. 2(a)-4(a)). Therefore, by including the $S'$ wave the $LF$ result becomes much closer to the data. However, it should be mentioned that only $\simeq 65 \%$ of the experimental value of the neutron charge radius can be at most reproduced by our $LF$ calculations, which, we remind, are based on the assumption of point-like $CQ$'s. Note that a non-vanishing $CQ$ size can affect the calculated neutron charge radius only if it is flavour dependent, i.e. different for $u$ and $d$ constituent quarks. The introduction of the effects of a possible non-vanishing $CQ$ size is suggested also by the overestimate of the experimental points for $Q^2 \gsim 0.5 ~ (GeV/c)^2$ (see Fig. 4(a)), but the estimate of such effects is beyond the aims of the present work.

\begin{figure}[htb]

\vspace{0.15cm}

\centerline{\epsfxsize=16cm \epsfig{file=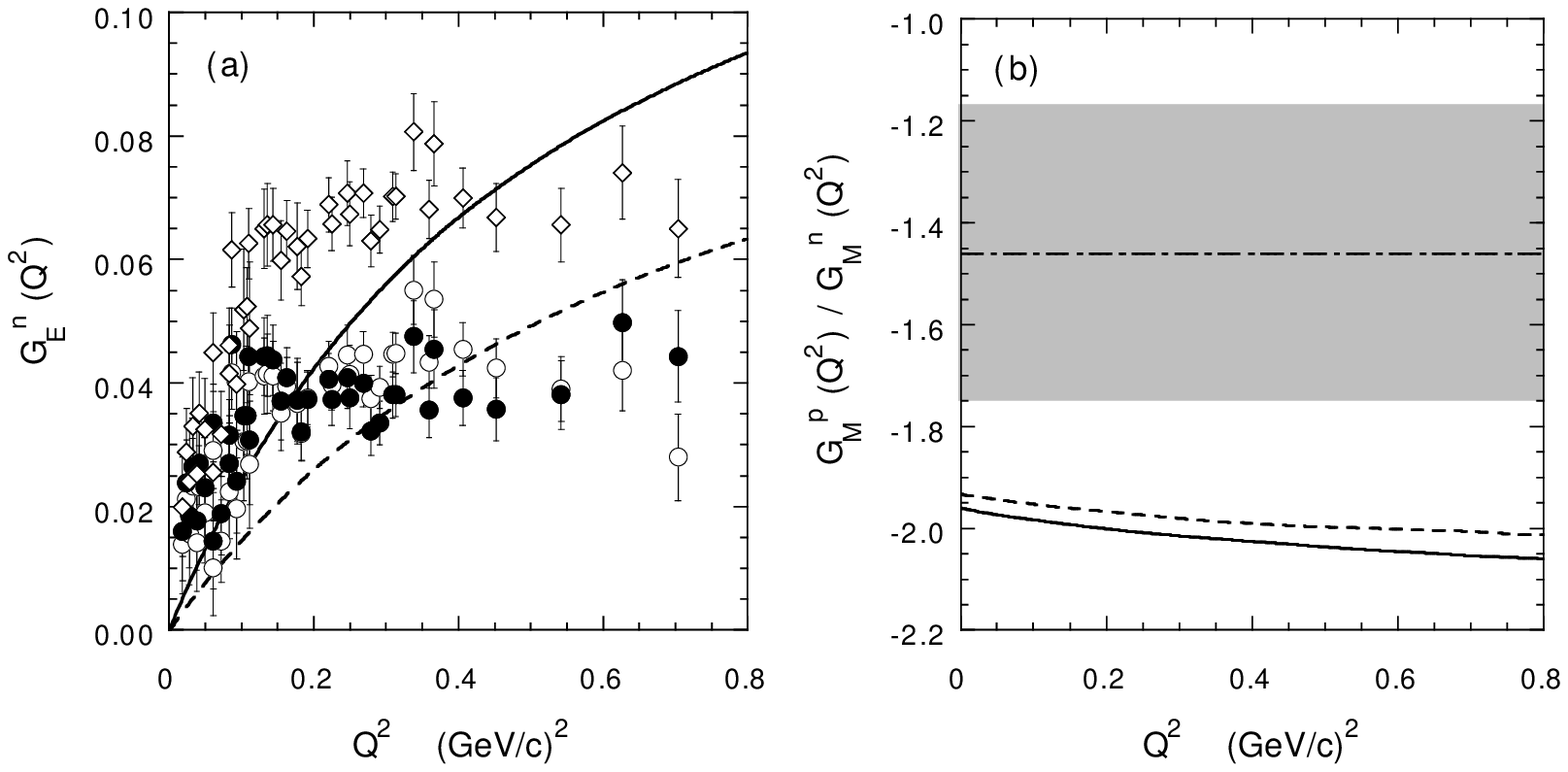}}

{\small {\bf Figure 4.} The same as in Fig. 2, but within the $LF$ approach (Eq. (\protect\ref{eq:F1F2_LF})) based on the use of the {\em plus} component of the one-body e.m. current only.}

\vspace{0.15cm}

\end{figure}

\indent As for $G_M^p(Q^2) / G_M^n(Q^2)$ (see Fig. 4(b)), the impact of the mixed-symmetry $S'$-wave is quite limited and therefore, at variance with the $NR$ limit and the $ZB$ approximation, the full $LF$ calculation is not consistent with the data on the magnetic form factor ratio. In the next Section we illustrate that such a failure results from spurious effects due to the loss of the rotational covariance in the $LF$ formalism, which occurs for approximate current operators, like the one-body e.m. current given by Eq. (\ref{eq:current}). On the contrary, we illustrate also that the calculations of the nucleon {\em charge} form factor $G_E^N(Q^2)$ based on the {\em plus} component of the e.m. current are free from unwanted spurious effects. To this end we will make use of the covariant $LF$ formalism, which has been recently reviewed in Ref. \cite{Carbonell}.

\section{The rotational covariance problem}

\indent As it is well known (cf., e.g., Ref. \cite{KP91}), in the $LF$ formalism the requirement of the full Poincar\'e covariance of the e.m. current operator is not fulfilled by the one-body current (\ref{eq:current}). This failure is related to the fact that the transverse rotations (with respect to the direction of the spin-quantization axis $\hat{n}$) cannot be kinematical and therefore depend upon the interaction.

\indent An explicit manifestation of the loss of the rotational covariance is the so-called angular condition. As already pointed out in Section 3, the physical form factors appearing in the covariant decomposition of a conserved current can be expressed in terms of the matrix elements of only one component of the current, namely the {\em plus} component. It may occur however that the number of physical form factors is less than the number of the independent matrix elements of the {\em plus} component, obtained from the application of general symmetry properties to the current operator. This means that in such situations a relation among the matrix elements (the so-called angular condition) should occur in order to constrain further their number. Within the $LF$ constituent quark model we have investigated two particular cases in Refs. \cite{CAR_rho} and \cite{CAR_Delta}, namely the elastic form factors for the $\rho$-meson and the $N - \Delta(1232)$ transition form factors. In both cases the angular condition can be formulated and we have shown that the use of the one-body current (\ref{eq:current}) lead to important violations of the angular condition, which can even totally forbid the extraction of the physical form factors from the matrix elements of the {\em plus} component of the current. This problem turns out to be particularly severe in case of "small" form factors, like e.g. the $E2 / M1$ ratio for the $N - \Delta(1232)$ transition \cite{CAR_Delta}.

\indent In case of the nucleon elastic form factors it is not possible to formulate an angular condition, because we have the same number of physical form factors [i.e., $F_1^N(Q^2)$ and $F_2^N(Q^2)$ appearing in the covariant decomposition (\ref{eq:F1F2})] and of independent matrix elements of the {\em plus} component of the current (see Eq. (\ref{eq:Iplus})). However, this does not mean that rotational covariance is fulfilled. Indeed, we now make manifest the loss of the rotational covariance in our $LF$ calculations by noting that the nucleon form factors can be extracted using not only the {\em plus} component of the e.m. current operator, but also through other components. As a matter of fact, adopting the Breit frame specified in Section 3, it is straightforward to get
 \be 
    \langle \Psi_{N}^{\nu'_N}| ~ {\cal{I}}^y ~ |\Psi_N^{\nu_N} \rangle =
    \left[ F_1^N(Q^2) + F_2^N(Q^2) \right] {iQ \over 2P^+} \langle \nu'_{N} 
     | \sigma_z | \nu_{N} \rangle
    \label{eq:Iy}
 \ee
 Therefore,  the nucleon magnetic form factor $G_M^N(Q^2)$ can be obtained from the matrix elements of the $y$-component of the e.m. current, where $y$ is the transverse axis orthogonal to $\vec{q}_{\perp}$, viz.
\be
    G_M^N(Q^2) = - {P^+ \over Q} \mbox{Tr}[{\cal{I}}^y i \sigma_z]
   \label{eq:GM_y}
\ee
For the {\em exact} e.m. current the use of Eq. (\ref{eq:F1F2+}) and Eq. (\ref{eq:GM_y}) should yield the same result for $G_M^N(Q^2)$. However, this would not be the case for an approximate current, like the one-body current (\ref{eq:current}) employed in our calculations. As a matter of fact, the results obtained for $G_M^N(Q^2)$ using  Eq. (\ref{eq:F1F2+}) and Eq. (\ref{eq:GM_y}) are reported in Fig. 5, where it can clearly be seen that different components of the one-body current lead to quite different results. This finding is a clear manifestation of the presence of spurious, unphysical effects related to the loss of the rotational covariance in the $LF$ formalism. It is interesting to note that we have carried out the calculations of $G_M^N(Q^2)$ using the $y$-component of the one-body e.m. current also within the $NR$ limit and the $ZB$ approximation. In both approximations the use of the {\em plus} and the $y$ components yield the same result, i.e. spurious effects are found to be completely absent. While this finding is quite obvious in the $NR$ limit (because the loss of the rotational covariance can occur only in relativistic approaches), the absence of spurious effects in the $ZB$ approximation implies that the loss of rotational covariance in the $LF$ formalism is related to the occurrence of the $CQ$ initial transverse motion. Note finally from Fig. 5 that the $LF$ calculations  closer to the results of the $NR$ limit are those based on the $y$-component.

\begin{figure}[htb]

\vspace{0.15cm}

\centerline{\epsfxsize=12cm \epsfig{file=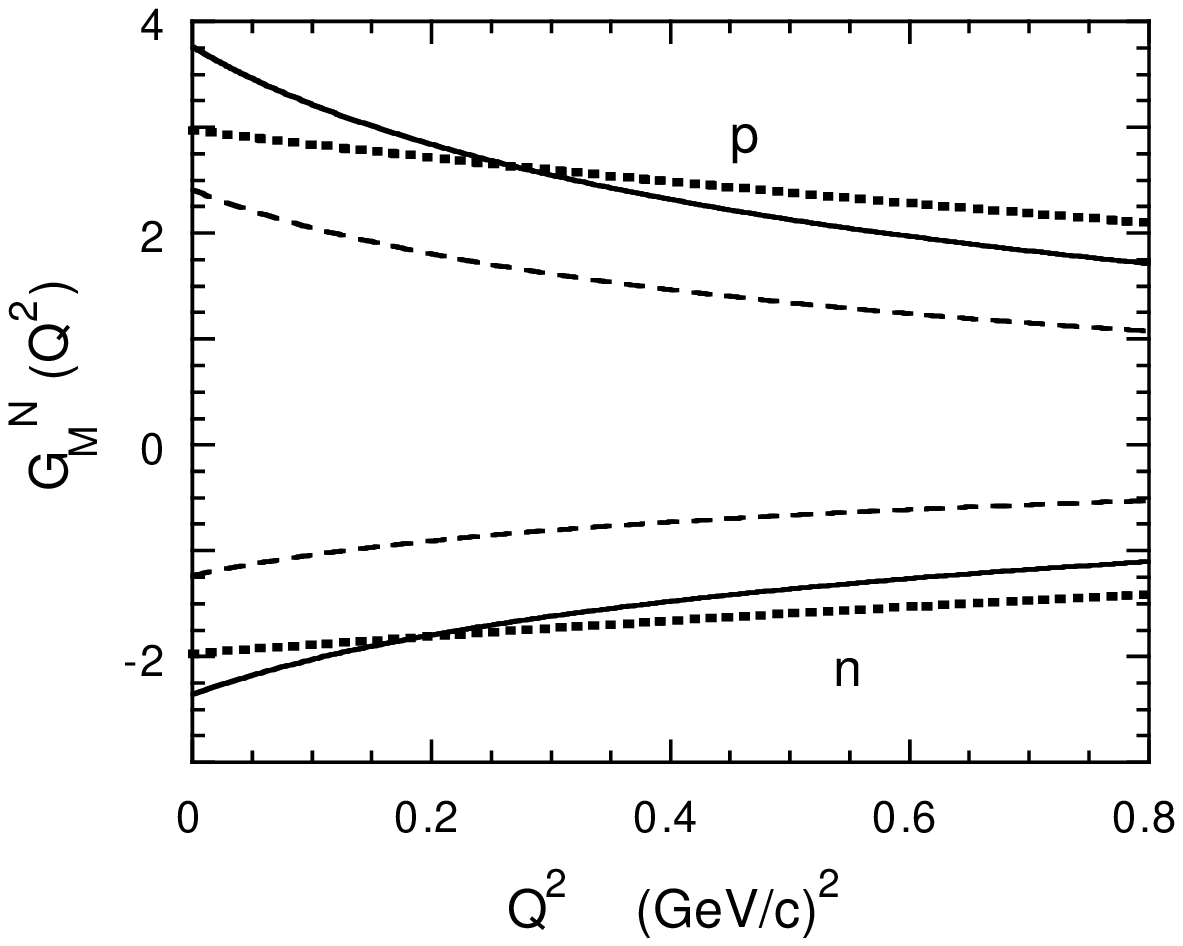}}

{\small {\bf Figure 5.} The nucleon magnetic form factor $G_M^N(Q^2)$ vs. $Q^2$. Dotted lines: NR limit (Eq. (\protect\ref{eq:GEGM_nr})). Dashed lines: results obtained from the {\em plus} component of the one-body e.m. current (Eq. (\protect\ref{eq:F1F2+})). Solid lines: results obtained using the $y$- component of the one-body e.m. current (Eq. (\protect\ref{eq:GM_y})). Upper and lower lines correspond to proton and neutron, respectively. Point-like $CQ$'s are assumed.}

\vspace{0.15cm}

\end{figure}

\indent A possible way to circumvent the above mentioned spurious effects has been developed in Ref. \cite{Karmanov}. There, it has been argued that the loss of the rotational covariance for an approximate current implies the dependence of its matrix elements upon the choice of the null four-vector $\omega \equiv (1, \hat{n})$ defining the spin-quantization axis. The particular direction defined by $\omega$ is irrelevant only if the current satisfies rotational covariance. Therefore, the covariant decomposition of an approximate current can still be done provided the four-vector $\omega$ is explicitly considered in the construction of the relevant covariant structures. In case of on-shell nucleons (i.e., $p^2 = p'^2 = M^2$), instead of Eq. (\ref{eq:F1F2}) one has \cite{Karmanov}
 \be
     <N(p',s')| \tilde{J}^{\mu}(0) |N(p,s)> & = & \bar{u}(p',s') \left \{ 
     F_1^N(Q^2) \gamma^{\mu} + F_2^N(Q^2) {i \sigma^{\mu \nu} q_{\nu} \over 
     2M} + \right. \nonumber \\[3mm]
     & & \left. B_1^N(Q^2) \left[ {\gamma \cdot \omega \over \omega \cdot P} 
     - {1 \over M (1 + \eta) } \right] P^{\mu} + B_2^N(Q^2) {\omega^{\mu} 
     \over \omega \cdot P} + \right. \nonumber \\[3mm]
     & & \left. B_3^N(Q^2) {\gamma \cdot \omega ~ \omega^{\mu} \over (\omega 
     \cdot P)^2} \right \} u(p,s)
     \label{eq:nucleon}
 \ee
where $F_1^N(Q^2)$ and $F_2^N(Q^2)$ are the physical form factors related to a generic approximate e.m. current operator $\tilde{J}$, whereas the form factors $B_i^N(Q^2)$ (with $i = 1, 2, 3$) are {\em unphysical} and multiply the covariant structures depending on $\omega$. Eventually, in Eq. (\ref{eq:nucleon}) $P = (p + p') /2$ and $\eta = Q^2 / 4 M^2$. Note that \cite{Karmanov}: ~ i) in the decomposition (\ref{eq:nucleon}) all possible gauge-dependent terms are forbidden because of time-reversal invariance; ~ ii) both the physical and unphysical form factors do not depend explicitly upon $\omega$, because we have chosen a reference frame where $q^+ = 0$, and ~ iii) for the exact current one must have $B_i^N(Q^2) = 0$.

\indent Considering the {\em plus} component in Eq. (\ref{eq:nucleon}) one gets
 \be 
    \langle \Psi_{N}^{\nu'_N}| ~ \tilde{J}^+ ~ |\Psi_N^{\nu_N} \rangle & = &
    \left[ F_1^N(Q^2) + {\eta \over 1 + \eta} B_1^N(Q^2) \right] 
    \delta_{{\nu'}_N \nu_{N}} - \nonumber \\[3mm]
    & & i {Q \over 2 M} \left[ F_2^N(Q^2) + {1 \over 
    1 + \eta} B_1^N(Q^2) \right] \langle \nu'_{N} | \sigma_y | \nu_{N} 
     \rangle
    \label{eq:Jplus}
 \ee
which implies that for an approximate e.m. current an equation similar to (\ref{eq:Iplus}) holds, but with the physical form factors $F_1^N(Q^2)$ and $F_2^N(Q^2)$ replaced by the following combinations
 \be
      \tilde{F}_1^N(Q^2) & = & F_1^N(Q^2)  + {\eta \over 1 + \eta} 
      B_1^N(Q^2)  ~ , \nonumber \\[3mm]
      \tilde{F}_2^N(Q^2) & = & F_2^N(Q^2)  + {1 \over 1 + \eta} B_1^N(Q^2)
      \label{eq:F1F2_tilde}
 \ee
containing the spurious form factor $B_1^N(Q^2)$. In terms of the Sachs form factors (\ref{eq:sachs}) one obtains
 \be
     \tilde{G}_E^N(Q^2) & \equiv & \tilde{F}_1^N(Q^2) - \eta 
     \tilde{F}_2^N(Q^2) = F_1^N(Q^2) - \eta F_2^N(Q^2) = G_E^N(Q^2) ~ , 
     \nonumber \\[3mm]
    \tilde{G}_M^N(Q^2) & \equiv & \tilde{F}_1^N(Q^2) + \tilde{F}_2^N(Q^2) = 
     F_1^N(Q^2) + F_2^N(Q^2) + B_1^N(Q^2) \neq G_M^N(Q^2)
     \label{eq:GEGM_tilde}
 \ee
Therefore, the nucleon charge form factor $G_E^N(Q^2)$ can be safely determined from the matrix elements of the {\em plus} component of the e.m. current, while the same is not true for the nucleon magnetic form factor $G_M^N(Q^2)$. However, from Eq. (\ref{eq:nucleon}) it follows that
 \be 
    \langle \Psi_{N}^{\nu'_N}| ~ \tilde{J}^y ~ |\Psi_N^{\nu_N} \rangle =
    \left[ F_1^N(Q^2) + F_2^N(Q^2) \right] {iQ \over 2P^+} \langle \nu'_{N} 
     | \sigma_z | \nu_{N} \rangle
    \label{eq:Jy}
 \ee
and therefore $G_M^N(Q^2)$ can be safely determined from the  matrix elements of the $y$-component of the e.m. current (as in Eq. (\ref{eq:GM_y})).

\indent In Fig. 6 we have reported our final results based on the use of Eq. (\ref{eq:F1F2+}) for $G_E^n(Q^2)$ and on Eq. (\ref{eq:GM_y}) for the ratio $G_M^p(Q^2) / G_M^n(Q^2)$. It can be seen that, provided spurious effects are properly avoided, our $LF$ results are fully consistent with the experimental data on $G_M^p(Q^2) / G_M^n(Q^2)$.

\begin{figure}[htb]

\vspace{0.15cm}

\centerline{\epsfxsize=16cm \epsfig{file=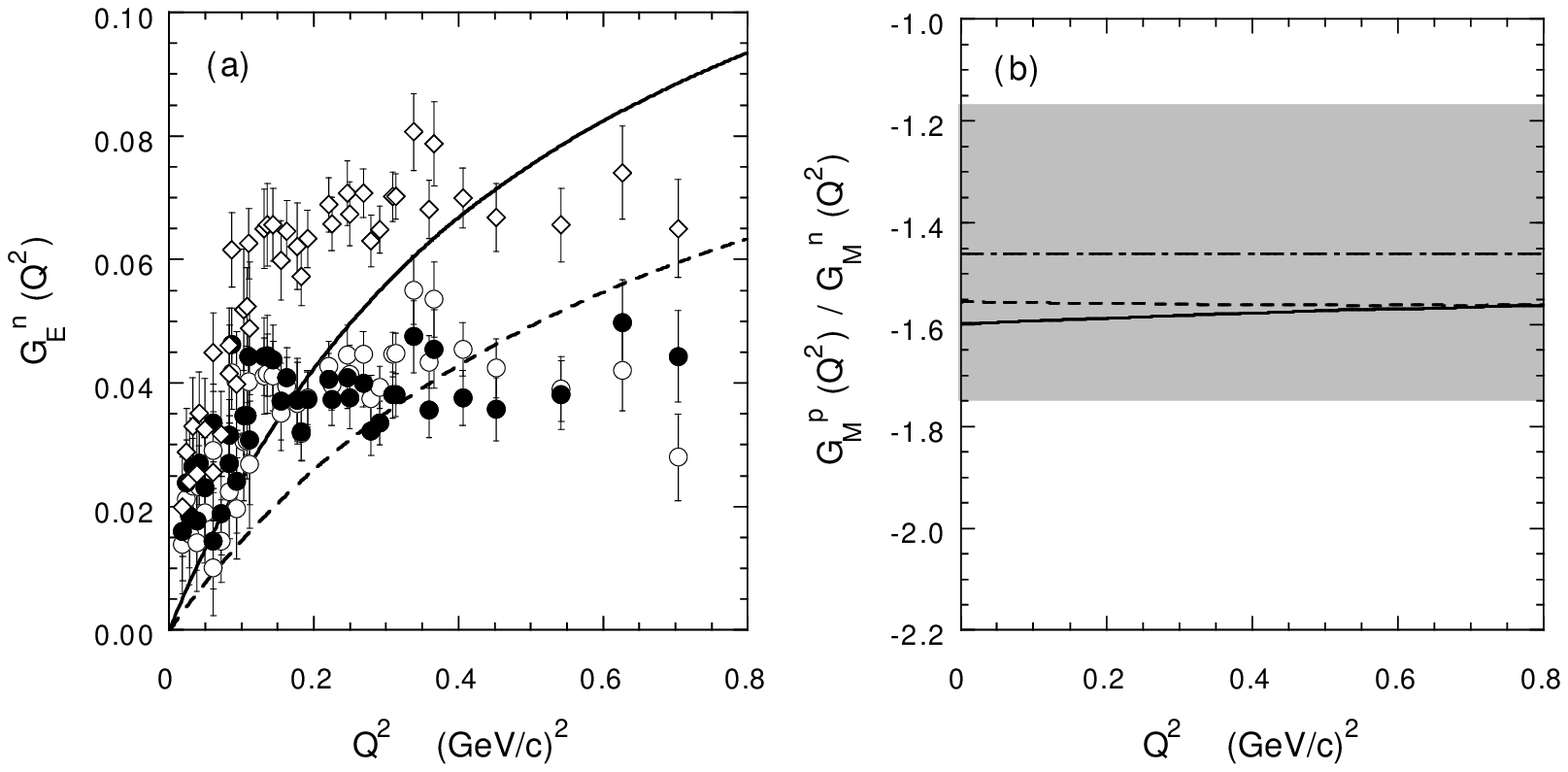}}

{\small {\bf Figure 6.} The same as in Fig. 2, but within the $LF$ approach using the {\em plus} component of the one-body e.m. current (see Eq. (\protect\ref{eq:F1F2+})) for the determination of $G_E^n(Q^2)$ and the transverse $y$-component (see Eq. (\protect\ref{eq:GM_y})) for the calculation of the nucleon magnetic form factors.}

\vspace{0.15cm}

\end{figure}

\indent To sum up, the most relevant $SU(6)$ breaking exhibited by the experimental data, namely $G_E^n(Q^2) \neq 0$ and $G_M^p(Q^2) / G_M^n(Q^2) \neq - 3/ 2$, can be understood qualitatively as well as quantitatively within the $CQ$ model, provided the effects of the spin-dependent components of the effective quark-quark interaction are taken into account as well as the relativistic effects arising from the $LF$ composition of the $CQ$ spins are properly considered. 

\section{The ratio $G_E^p(Q^2) / G_M^p(Q^2)$}

\indent In this Section we will briefly address the issue of the interpretation of the recently observed \cite{JLAB} deviation of the ratio $\mu_p G_E^p(Q^2) / G_M^p(Q^2)$ from the dipole-fit expectation $\mu_p G_E^p(Q^2) /$ $G_M^p(Q^2) = 1$, where $\mu_p = G_M^p(0)$ is the proton magnetic moment. First of all let us note that if the mixed-symmetry $S'$-wave is neglected and point-like $CQ$'s are assumed, the $NR$ limit (\ref{eq:GEGM_nr}) predicts a ratio $\mu_p G_E^p(Q^2) / G_M^p(Q^2) = 1$ (i.e., coinciding with the dipole-fit expectation),  while the $ZB$ approximation (\ref{eq:GEGM_zb}) yields $\mu_p G_E^p(Q^2) / G_M^p(Q^2) = 1 - Q^2 / 2M^2$, where now $\mu_p$ is the calculated proton magnetic moment.

\begin{figure}[htb]

\vspace{0.15cm}

\centerline{\epsfxsize=16cm \epsfig{file=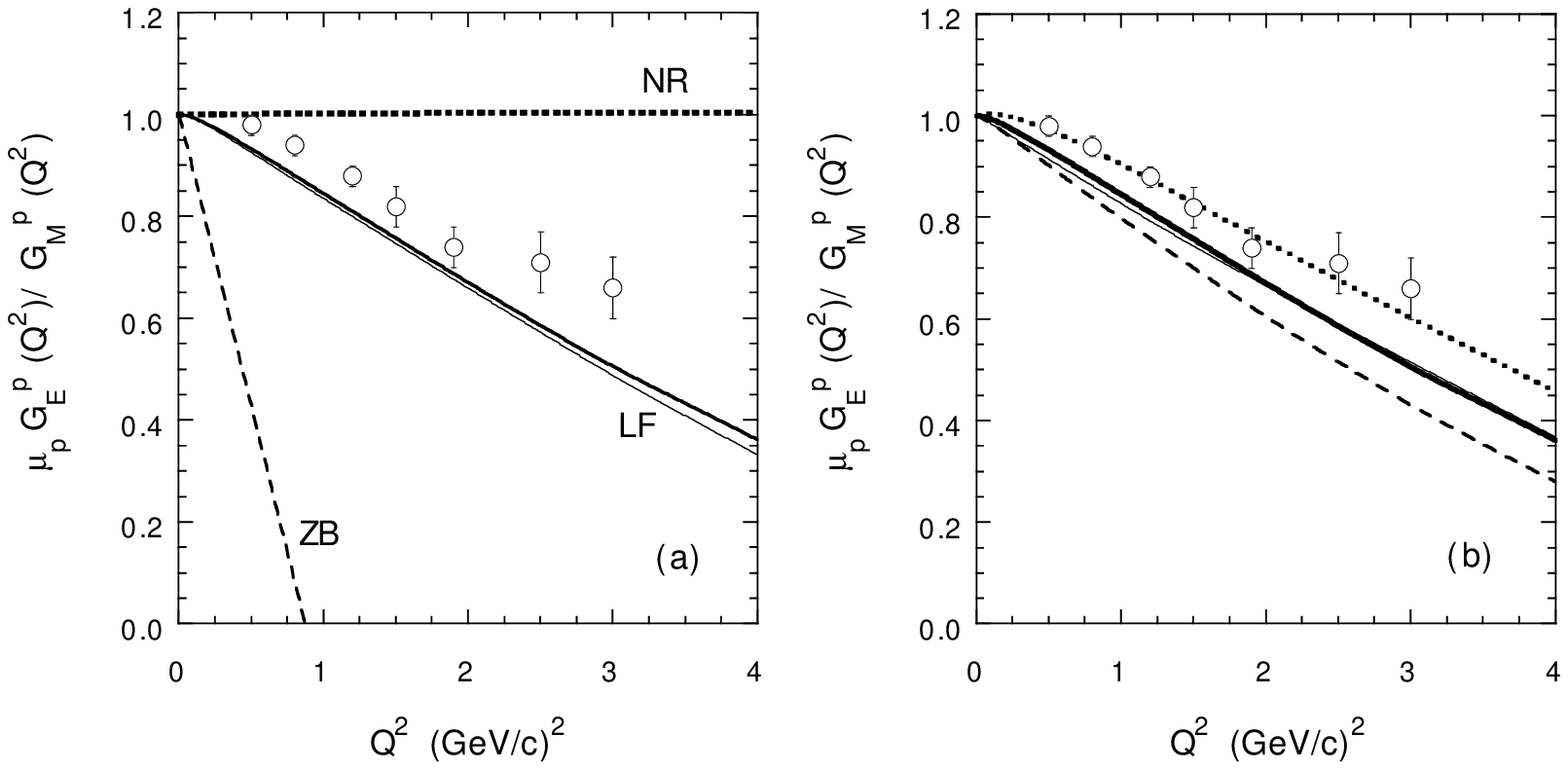}}

{\small {\bf Figure 7.} Ratio $\mu_p G_E^p(Q^2) / G_M^p(Q^2)$ versus $Q^2$. Open squares are the recent experimental points from $JLAB$ \protect\cite{JLAB}. (a) Dotted, dashed and solid lines correspond to the predictions of the $NR$ limit (\protect\ref{eq:GEGM_nr}) and the $ZB$ approximation (\protect\ref{eq:GEGM_zb}), as well as to our full $LF$ calculations [namely Eq. (\protect\ref{eq:F1F2+}) for $G_E^p(Q^2)$ and Eq. (\protect\ref{eq:GM_y}) for $G_M^p(Q^2)$], respectively. Thick and thin lines are the results obtained with and without the effects from the mixed-symmetry $S'$-wave (dotted and dashed thin lines are undistinguishable from the thick ones), respectively. (b) The thick solid line is the result obtained using the full $OGE$ interaction model \cite{CI86} (yielding $\langle p_{\perp} \rangle = 0.58 ~ GeV$), while the dashed line corresponds to the case of its linear confinement term only (corresponding to $\langle p_{\perp} \rangle = 0.33 ~ GeV$). The thin solid line corresponds to the use of the chiral quark potential model of Ref. \cite{Glozmann} (having $\langle p_{\perp} \rangle = 0.61 ~ GeV$), while the dotted line is the result obtained adopting a pure $SU(6)$-symmetric $HO$ ans\"atz corresponding to $\langle p_{\perp} \rangle = 1.0 ~ GeV$, respectively. In all the calculations point-like $CQ$'s are assumed.}

\vspace{0.15cm}

\end{figure}

\indent The full $LF$ calculations, based on  Eq. (\ref{eq:F1F2+}) for $G_E^p(Q^2)$ and Eq. (\ref{eq:GM_y}) for $G_M^p(Q^2)$, and the results obtained in the $NR$ limit and the $ZB$ approximation are reported in Fig. 7(a) and compared with the recent $JLAB$ data \cite{JLAB}. It can be seen that: ~ i) the impact of the mixed-symmetry $S'$-wave is very limited in each approach, and ~ ii) the predictions of both the $NR$ limit and the $ZB$ approximation are completely at variance with the $JLAB$ data; in particular, the $ZB$ approximation predicts a negative value for $\mu_p G_E^p(Q^2) / G_M^p(Q^2)$  for $Q^2 > 2M^2$ (with $M = 3 \cdot 0.220 ~ GeV$ for the $NR$ and $ZB$ calculations); this clearly signals that the applicability of the $ZB$ approximation should be limited only to low values of $Q^2$, namely $Q^2 << M^2$ (cf. also Ref. \cite{SIM99}). To sum up, our findings imply that a suppression of the ratio $\mu_p G_E^p(Q^2) / G_M^p(Q^2)$ from unity can be expected in the $CQ$ model provided the relativistic effects generated by the Melosh rotations of the $CQ$ spins are taken into account.

\indent The $LF$ results obtained adopting the $OGE$ and chiral quark potential models are reported in Fig. 7(b) and compared with calculations corresponding to various values of the average $CQ$ transverse momentum $\langle p_{\perp} \rangle$. It can clearly be seen that the calculated suppression of the ratio $\mu_p G_E^p(Q^2) / G_M^p(Q^2)$ from unity exhibits a slight dependence on the value of $\langle p_{\perp} \rangle$, and moreover the results corresponding to the $OGE$ and chiral quark potential models are very similar and compare quite favourably against the recent $JLAB$ data. A nice reproduction of the latter appears to be achieved when $\langle p_{\perp} \rangle \simeq 1 ~ GeV$ [see dotted line in Fig. 7(b)], but it should be reminded that our results have been obtained assuming point-like $CQ$'s, while the introduction of $CQ$ form factors is expected to affect (at least partially) the ratio $\mu_p G_E^p(Q^2) / G_M^p(Q^2)$.

\section{Conclusions}

\indent In this contribution we have investigated the effects of both kinematical and dynamical $SU(6)$ breaking on the nucleon elastic electromagnetic form factors within the constituent quark model formulated on the light-front. We have shown that the most relevant $SU(6)$ breaking exhibited by the experimental data, namely $G_E^n(Q^2) \neq 0$ and $G_M^p(Q^2) / G_M^n(Q^2) \neq - 3/ 2$, can be understood qualitatively as well as quantitatively within the constituent quark model, provided the effects of the spin-dependent components of the quark-quark interaction are taken into account as well as the relativistic effects arising from the light-front composition of the constituent spins are properly considered. We have indeed shown that the evaluation of the nucleon magnetic form factors has to be performed using the transverse $y$-component of the e.m. current in order to avoid spurious, unphysical effects related to the loss of the rotational covariance in the light-front formalism, while the nucleon charge form factor can be safely extracted from the matrix elements of the {\em plus} component of the current.

\indent  Finally, we have shown that a suppression of the ratio $G_E^p(Q^2) / G_M^p(Q^2)$ with respect to the dipole-fit prediction can be expected in the constituent quark model provided the relativistic effects generated by the Melosh rotations of the constituent spins are taken into account. The strength of the suppression exhibits a slight dependence on the value of the average quark transverse momentum in the nucleon; moreover, the results of the calculations based on the nucleon wave functions arising from two of the most sophisticated quark potential models \cite{CI86,Glozmann}, compare quite favourably against the recent data  \cite{JLAB} from Jefferson Lab.

\section*{Acknowledgments}

\indent The authors gratefully acknowledge V.A. Karmanov for many fruitful and valuable discussions on the covariant light-front formalism.\\

\noindent The authors want to dedicate this work to Nimai C. Mukhopadhyay.

\newpage

\section*{Appendix}

\subsection*{A. \parbox[t]{16cm}{Expansion of the wave function in the harmonic oscillator basis having definite permutational symmetry}}

The wave functions $w_S(\vec{k}, \vec{p})$, $w_{S'_s}(\vec{k}, \vec{p})$, $w_{S'_a}(\vec{k}, \vec{p})$ and $w_A(\vec{k}, \vec{p})$, appearing in Eq. (\ref{eq:canonical}), can be constructed by expanding them onto the complete harmonic oscillator ($HO$) basis, viz.
 \be
      w_{[f]}(\vec{k}, \vec{p}) = \sum_{\tilde{\rho}} C_{\tilde{\rho}}^{[f]} 
      ~ \psi_{\tilde{\rho}, 00}^{(HO, [f])}(\vec{k}, \vec{p}) 
      \label{eq: expansion}
 \ee
where $\psi_{\tilde{\rho}, L M_L}^{(HO, [f])}(\vec{k}, \vec{p})$ is a function of the $HO$ basis corresponding to total orbital angular momentum $L$ and its projection $M_L$, $\tilde{\rho}$ stands for all the other quantum numbers necessary to identify the basis functions (see later on) and $[f]$ identifies the component of the nucleon wave function (i.e., $[f] = S, S'_s, S'_a, A$). Finally, the quantities $C_{\tilde{\rho}}^{[f]}$ are variational coefficients, that can be determined by applying the Raleigh-Ritz variational principle to the Hamiltonian of the three-quark system.

\indent The basis function $\psi_{\tilde{\rho}, L M_L}^{(HO, [f])}(\vec{k}, \vec{p})$ can be written in the form
 \be
      \psi_{\tilde{\rho}, L M_L}^{(HO, [f])}(\vec{k}, \vec{p}) = \sum_{\rho} 
      ~  U_{\rho \tilde{\rho}}^{L, [f]} ~ \phi_{\rho, L 
      M_L}^{(HO,)}(\vec{k}, \vec{p})
     \label{eq:U[f]}
 \ee
where $\rho$ is a shorthand notation for the $HO$ radial and orbital quantum numbers $\rho = \{ n_k, \ell_k, n_p, \ell_p \}$ and $\phi_{\rho, L M_L}^{(HO,)}(\vec{k}, \vec{p})$ is the usual $HO$ wave function given explicitly by
 \be
      \phi_{\rho, L M_L}^{(HO)}(\vec{k}, \vec{p}) = R_{n_k 
      \ell_k}^{(HO)}(k) ~ R_{n_p \ell_p}^{(HO)}(p) ~ \sum_{m_k m_p} \langle 
      \ell_k m_k, \ell_p, m_p | L M_L \rangle Y_{\ell_k m_k}(\hat{k}) 
      Y_{\ell_p m_p}(\hat{p})
      \label{eq:HOwf}
 \ee
with $R_{n_k \ell_k}^{(HO)}(k)$ and $R_{n_p \ell_p}^{(HO)}(p)$ being the radial $HO$ functions depending on the $HO$ length $a_{HO}$ [namely, taking into account the definitions (\ref{eq:jacobian}), one has: $R_{n_k \ell_k}^{(HO)}(k) \propto exp(- k^2 / 2 a_{HO}^2)$ and $R_{n_p \ell_p}^{(HO)}(p) \propto exp(- 3 p^2 / 8 a_{HO}^2)$]. Thanks to the use of only one $HO$ length, in Eq. (\ref{eq:U[f]}) the sum over $\rho$ is limited to the $HO$ functions having the same number of $HO$ excitation quanta $N_{HO} \equiv N_k + N_p = 2n_k \ell_k + 2n_p +\ell_p$, the same orbital parity ${\bf{\pi}} \equiv (-)^{\ell_k + \ell_p}$ and the same orbital angular momentum $L$ and its projection $M_L$. In Eq. (\ref{eq:U[f]}) the matrix $U_{\rho \tilde{\rho}}^{L, [f]}$, which is independent on $M_L$ thanks to the Wigner-Eckart theorem, is explicitly given by
 \be
      U_{\rho \tilde{\rho}}^{L, [f]=S} & = & \sqrt{{2 \over 1 + 
     \delta_{\tilde{n}_k \tilde{n}_p} \delta_{\tilde{\ell}_k 
     \tilde{\ell}_p}}} {1 + (-)^{\ell_k} \over 2} (-)^{n_k + {\ell_k \over 
     2} + \ell_p} \langle \rho; L | \tilde{\rho}; L \rangle \Delta_{d, 0} ~ 
      , \nonumber \\[3mm]
      U_{\rho \tilde{\rho}}^{L, [f]=S'_s} & = & \sqrt{2} {1 + (-)^{\ell_k} 
      \over  2} (-)^{n_k + {\ell_k \over 2} + \ell_p} \langle \rho; L | 
      \tilde{\rho}; L \rangle \left[ \Delta_{d, 1} + \Delta_{d, 2} \right] ~ 
      , \nonumber \\[3mm]
      U_{\rho \tilde{\rho}}^{L, [f]=S'_a} & = & \sqrt{2} {1 - (-)^{\ell_k} 
      \over 2} (-)^{n_k + {\ell_k + 1 \over 2} + \ell_p} \langle \rho; L | 
      \tilde{\rho}; L \rangle \left[ -\Delta_{d, 1} + \Delta_{d, 2} \right] 
      ~ , \nonumber \\[3mm]
      U_{\rho \tilde{\rho}}^{L, [f]=A} & = & \sqrt{2} {1 - (-)^{\ell_k} 
      \over 2} (-)^{n_k + {\ell_k + 1 \over 2} + \ell_p} \langle \rho; L | 
      \tilde{\rho}; L \rangle \Delta_{d, 0}
      \label{eq:S3}
 \ee
where $\langle \rho; L | \tilde{\rho}; L \rangle$ is a shorthand notation for the Brody-Moshinsky coefficients \cite{Moshinsky}, $d \equiv N_k - N_p = 2n_k + \ell_k - 2n_ p - \ell_p$, and $\Delta_{d, m} = 1$ if $d = m$ [$\mbox{mod}(3)$] while $\Delta_{d, m} = 0$ otherwise. Finally, the index $\tilde{\rho}$ stands for $\tilde{\rho} = \{ \tilde{n}_k, \tilde{\ell}_k, \tilde{n}_p, \tilde{\ell}_p \}$, but the following constraints apply: ~ i) $\tilde{n}_k \geq \tilde{n}_p$, ~ ii) $\tilde{\ell}_k \geq \tilde{\ell}_p$ if $\tilde{n}_k = \tilde{n}_p$, and ~ iii) $\tilde{n}_k \neq \tilde{n}_p$ for $[f] = A$ and $[f] = S$ if $L$ is odd.

\indent Note that in the right-hand sides of Eq. (\ref{eq:S3}) the presence of the factor $(-)^{\ell_p}$ is related to the definition of the jacobian variable $\vec{p}$ as given in Eq. (\ref{eq:jacobian}).

\indent In the calculations presented in this work we have employed the $HO$ basis up to $20$ $HO$ excitation quanta (i.e. $N_{HO} \leq 20$), which corresponds to a total of $67$ basis states for the symmetric $S$-wave, $94$ states for the mixed-symmetry $S'$-wave and $31$ states for the antisymmetric $A$-wave. By combining these configuration states with the proper spin-isospin functions, one gets a total of $192$ (completely symmetric) $HO$ basis states for the expansion of the (canonical) nucleon wave function (\ref{eq:canonical}). Note that the number of $HO$ functions $\phi_{\rho, L M_L}^{(HO,)}(\vec{k}, \vec{p})$, which appear in Eq. (\ref{eq:U[f]}) and do not posses any definite permutational symmetry, is $572$ for $N_{HO} \leq 20$ and $L = M_L =0$.

\subsection*{B. Evaluation of the coefficients ${\cal{R}}_{S'_{12} S_{12}}^{(\alpha \beta)}$}

\indent In Eq. (\ref{eq:F1F2_LF}) the coefficients ${\cal{R}}_{S'_{12} S_{12}}^{(\alpha \beta)}$ contains the effects of the Melosh rotations of the $CQ$ spins and are defined as
 \be
     {\cal{R}}_{S'_{12} S_{12}}^{(\alpha=1 ~ \beta)} & = & {1 \over 2} 
     \sum_{\nu_N} \langle \left( S'_{12} ~ {1 \over 2} \right) {1 \over 2} 
     \nu_N |  ~ {\cal{R}}^{\dag} ~ O^{\beta} ~ {\cal{R}} ~ | \left( S_{12} ~ 
     {1 \over 2} \right) {1 \over 2} \nu_N \rangle \nonumber \\[3mm]
     {\cal{R}}_{S'_{12} S_{12}}^{(\alpha=2 ~ \beta)} & = & - {i \over 2} 
     \sum_{\nu_N \nu'_N}  \langle \left( S'_{12} ~ {1 \over 2} \right) {1 
     \over 2} \nu_N | ~ {\cal{R}}^{\dag} ~ O^{\beta} ~ {\cal{R}} ~ |  \left( 
     S_{12} ~ {1 \over 2} \right) {1 \over 2} \nu'_N \rangle ~ \langle {1 
     \over 2} \nu'_N |  ~ \sigma_y ~ |  {1 \over 2} \nu_N \rangle
     \label{eq:Ralphabeta}
 \ee
where $\cal{R}^{\dag}$ is given by Eqs. (\ref{eq:Rmelosh}-\ref{eq:melosh}), $O^1$ is the identity $2 \times 2$ matrix and $O^2 = -i \sigma_y$ acting on the spin of the particle $3$, and
 \be
     | \left( S_{12} ~ {1 \over 2} \right) {1 \over 2} \nu_N \rangle = 
     \sum_{M_S} \langle {1 \over 2} \nu_1 {1 \over 2} \nu_2 | S_{12} M_S 
     \rangle ~ \langle S_{12} M_S {1 \over 2} \nu_3 | {1 \over 2} \nu_N 
     \rangle
     \label{eq:spinwf}
 \ee
The sums over the spin projections yield
 \be
     {\cal{R}}_{00}^{(11)} & = & {\cal{N}} \left[ A_1 A_2 + \vec{B}_1 \cdot 
     \vec{B}_2 \right] A_3 ~ , \nonumber \\[3mm]
     {\cal{R}}_{00}^{(12)} & = & {\cal{N}} \left[ A_1 A_2 + \vec{B}_1 \cdot 
     \vec{B}_2 \right] \tilde{A}_3 ~ , \nonumber \\[3mm]
     {\cal{R}}_{00}^{(21)} & = & {\cal{N}} \left[ A_1 A_2 + \vec{B}_1 \cdot 
     \vec{B}_2 \right] B_3 ~ , \nonumber \\[3mm]
     {\cal{R}}_{00}^{(22)} & = & {\cal{N}} \left[ A_1 A_2 + \vec{B}_1 \cdot 
     \vec{B}_2 \right] \tilde{B}_3 ~ ,
    \label{eq:R00}
 \ee
 \be
      {\cal{R}}_{01}^{(11)} & = & - { 1 \over \sqrt{3}} {\cal{N}} \left[ A_2 
      \vec{B}_1 - A_1 \vec{B}_2 - \vec{B}_1 \times \vec{B}_2 \right] \cdot 
     \vec{B}_3 ~ , \nonumber \\[3mm]
      {\cal{R}}_{01}^{(12)} & = & - { 1 \over \sqrt{3}} {\cal{N}} \left[ A_2 
      \vec{B}_1 - A_1 \vec{B}_2 - \vec{B}_1 \times \vec{B}_2 \right] \cdot 
      \vec{\tilde{B}}_3 ~ , \nonumber \\[3mm]
      {\cal{R}}_{01}^{(21)} & = & { 1 \over \sqrt{3}} {\cal{N}} \left[ 
      \left( A_2 B_{1y} - A_1 B_{2y} + B_{2z} B_{1x} - B_{2x} B_{1z} \right) 
      A_3 + \right. \nonumber \\[3mm]
      & & \left. \left( A_1 B_{2x} - A_2 B_{1x} - B_{2y} B_{1z} + B_{2z} 
      B_{1y} \right) B_{3z} - \right. \nonumber \\[3mm]
      & & \left. \left( A_1 B_{2z} - A_2 B_{1z} - B_{2x} B_{1y} + B_{2y} 
      B_{1x} \right) B_{3x} \right] ~ , \nonumber \\[3mm]
      {\cal{R}}_{01}^{(22)} & = & { 1 \over \sqrt{3}} {\cal{N}} \left[ 
      \left( A_2 B_{1y} - A_1 B_{2y} + B_{2z} B_{1x} - B_{2x} B_{1z} \right) 
      \tilde{A}_3 + \right. \nonumber \\[3mm]
      & & \left.  \left( A_1 B_{2x} - A_2 B_{1x} - B_{2y} B_{1z} + B_{2z} 
      B_{1y} \right) \tilde{B}_{3z} -  \right. \nonumber \\[3mm]
      & & \left. \left( A_1 B_{2z} - A_2 B_{1z} - B_{2x} B_{1y} + B_{2y} 
      B_{1x} \right) \tilde{B}_{3x} \right] ~ , 
     \label{eq:R01}
 \ee
 \be
      {\cal{R}}_{10}^{(11)} & = & - { 1 \over \sqrt{3}} {\cal{N}} \left[ A_2 
      \vec{B}_1 - A_1 \vec{B}_2 + \vec{B}_1 \times \vec{B}_2 \right] \cdot 
     \vec{B}_3 ~ , \nonumber \\[3mm]
      {\cal{R}}_{10}^{(12)} & = & - { 1 \over \sqrt{3}} {\cal{N}} \left[ A_2 
      \vec{B}_1 - A_1 \vec{B}_2 + \vec{B}_1 \times \vec{B}_2 \right] \cdot 
      \vec{\tilde{B}}_3 ~ , \nonumber \\[3mm]
      {\cal{R}}_{10}^{(21)} & = & { 1 \over \sqrt{3}} {\cal{N}} \left[ 
      \left( A_2 B_{1y} - A_1 B_{2y} - B_{2z} B_{1x} + B_{2x} B_{1z} \right) 
      A_3 - \right. \nonumber \\[3mm]
      & & \left. \left( A_1 B_{2x} - A_2 B_{1x} + B_{2y} B_{1z} - B_{2z} 
      B_{1y} \right) B_{3z} + \right. \nonumber \\[3mm]
      & & \left. \left( A_1 B_{2z} - A_2 B_{1z} + B_{2x} B_{1y} - B_{2y} 
      B_{1x} \right) B_{3x} \right] ~ , \nonumber \\[3mm]
      {\cal{R}}_{10}^{(22)} & = & { 1 \over \sqrt{3}} {\cal{N}} \left[ 
      \left( A_2 B_{1y} - A_1 B_{2y} - B_{2z} B_{1x} + B_{2x} B_{1z} \right) 
      \tilde{A}_3 - \right. \nonumber \\[3mm]
      & & \left.  \left( A_1 B_{2x} - A_2 B_{1x} + B_{2y} B_{1z} - B_{2z} 
      B_{1y} \right) \tilde{B}_{3z} +  \right. \nonumber \\[3mm]
      & & \left. \left( A_1 B_{2z} - A_2 B_{1z} + B_{2x} B_{1y} - B_{2y} 
      B_{1x} \right) \tilde{B}_{3x} \right] ~ , 
     \label{eq:R10}
 \ee
 \be
     {\cal{R}}_{11}^{(11)} & = & {1 \over 3} {\cal{N}} \left[ \left( 3 A_1 
     A_2 - \vec{B}_1 \cdot \vec{B}_2 \right) A_3 + 2 \left( A_1 \vec{B}_2 + 
     A_2 \vec{B}_1 \right) \cdot \vec{B}_3 \right] ~ , \nonumber \\[3mm]
     {\cal{R}}_{11}^{(12)} & = & {1 \over 3} {\cal{N}} \left[ \left( 3 A_1 
     A_2 - \vec{B}_1 \cdot \vec{B}_2 \right) \tilde{A}_3 + 2 \left( A_1 
     \vec{B}_2 + A_2 \vec{B}_1 \right) \cdot \vec{\tilde{B}}_3 \right] ~ , 
     \nonumber \\[3mm]
     {\cal{R}}_{11}^{(21)} & = & {1 \over 3} {\cal{N}} \left[ 2 \left( 
     B_{1y} \vec{B}_2 + B_{2y} \vec{B}_1 \right) \cdot \vec{B}_3 - \left( 
     A_1 A_2 + \vec{B}_1 \cdot \vec{B}_2 \right) B_{3y} + \right.
     \nonumber \\[3mm]
     & & \left. 2 \left( A_1 B_{2y} + A_2 B_{1y} \right) A_3 \right] ~ , 
     \nonumber \\[3mm]
     {\cal{R}}_{11}^{(22)} & = & {1 \over 3} {\cal{N}} \left[ 2 \left( 
      B_{1y} \vec{B}_2 + B_{2y} \vec{B}_1 \right) \cdot \vec{\tilde{B}}_3 - 
     \left( A_1 A_2 + \vec{B}_1 \cdot \vec{B}_2 \right) \tilde{B}_{3y} + 
     \right. \nonumber \\[3mm]
     & & \left. 2 \left( A_1 B_{2y} + A_2 B_{1y} \right) \tilde{A}_3 \right] 
     ~ , 
    \label{eq:R11}
 \ee
where
 \be
    {\cal{N}} & = & \prod_{j = 1}^3 {1 \over \sqrt{(m + \xi_j M_0)^2 + k_{j 
    \perp}^2}} \cdot {1 \over \sqrt{(m + \xi_j {M'}_0)^2 + {k'}_{j 
    \perp}^2}} ~ , \nonumber \\[3mm]
    A_j & = & (m + \xi_j {M'}_0) ~ (m + \xi_j M_0) + \vec{k'}_{j \perp} 
    \cdot \vec{k}_{j \perp} ~ , \nonumber \\[3mm]
    \tilde{A}_3 & = & (m + \xi_3 {M'}_0) k_{3x} - (m + \xi_3 M_0) {k'}_{3x} 
    ~ , \nonumber \\[3mm] 
    B_{jx} & = & (m + \xi_j M_0) {k'}_{jy} - (m + \xi_j {M'}_0) k_{jy} ~ ,
    \nonumber \\[3mm]
    B_{jy} & = & (m + \xi_j {M'}_0) k_{jx} - (m + \xi_j M_0) {k'}_{jx} ~ ,
    \nonumber \\[3mm]
    B_{jz} & = & {k'}_{jx} k_{jy} - {k'}_{jy} k_{jx} ~ , \nonumber \\[3mm]
    \tilde{B}_{3x} & = & {k'}_{3x} k_{3y} + {k'}_{3y} k_{3x} ~ ,
    \nonumber \\[3mm]
    \tilde{B}_{3y} & = & - (m + \xi_3 {M'}_0) (m + \xi_3 M_0) - {k'}_{3x} 
    k_{3x} + {k'}_{3y} k_{3y} ~ ,  \nonumber \\[3mm]
    \tilde{B}_{3z} & = & (m + \xi_3 {M'}_0) k_{3y} + (m + \xi_3 M_0) 
    {k'}_{3y}
    \label{eq:ABN}
 \ee

\indent The $ZB$ approximation of Ref. \cite{Isgur} (see Eq. (\ref{eq:GEGM_zb})) can be obtained from the full $LF$ results (\ref{eq:F1F2_LF},\ref{eq:R00}-\ref{eq:R11}) making the assumptions explained in Ref. \cite{SIM99}. One easily obtains: ${\cal{R}}_{01}^{(\alpha \beta)} = {\cal{R}}_{10}^{(\alpha \beta)} = 0$ and
 \be
      {\cal{R}}_{00}^{(11)} & = & {m \over \sqrt{m^2 + Q^2 / 9}} ~ , 
     \nonumber \\[3mm] 
     {\cal{R}}_{00}^{(12)} & = & {1 \over 3} {Q \over 2m} {m \over \sqrt{m^2 
     + Q^2 / 9}} ~ , 
     \nonumber \\[3mm]
     {\cal{R}}_{00}^{(21)} & = & - {2 \over 3} {Q \over 2m} {m \over 
     \sqrt{m^2 + Q^2 / 9}} ~ , 
     \nonumber \\[3mm]
     {\cal{R}}_{00}^{(22)} & = & - {m \over \sqrt{m^2 + Q^2 / 9}} ~ , 
    \label{eq:R00_zb}
 \ee
 \be
      {\cal{R}}_{11}^{(11)} & = & {4m^2 - 3 Q^2 / 9 \over 4m^2 + Q^2 / 9} {m 
      \over \sqrt{m^2 + Q^2 / 9}} ~ , 
     \nonumber \\[3mm] 
     {\cal{R}}_{11}^{(12)} & = & - {1 \over 9} {Q \over 2m} {m \over 
     \sqrt{m^2 + Q^2 / 9}} ~ , 
     \nonumber \\[3mm]
     {\cal{R}}_{11}^{(21)} & = & {2 \over 3} {Q \over 2m} {4m^2 - Q^2 / 9 
     \over 4m^2 + Q^2 / 9} {m \over \sqrt{m^2 + Q^2 / 9}} ~ , 
     \nonumber \\[3mm]
     {\cal{R}}_{11}^{(22)} & = & {1 \over 3} {m \over \sqrt{m^2 + Q^2 / 9}} 
     ~ . 
    \label{eq:R11_zb}
 \ee


\newpage

\begin{thebibliography}{99}

\bibitem{Petratos} See, for a recent review, G. Petratos: Nucl. Phys.
 {\bf A666} (2000) 61.

\bibitem{Sachs} R.G. Sachs: Phys. Rev. {\bf 126} (1962) 2256.

\bibitem{SIM99} F. Cardarelli and S. Simula: Phys. Lett. {\bf B467} (1999) 
 1.

\bibitem{radius_n} S. Kopecki {\em et al.}: Phys. Rev. Lett. {\bf 74} (1995) 
 2427.

\bibitem{Weise} See, e.g., T. Ericson and W. Weise: {\em Pions and 
 Nuclei}, Claredon Press (Oxford, 1988).

\bibitem{Isgur} N. Isgur: Phys. Rev. Lett. {\bf 83} (1999) 272.

\bibitem{CI86} S. Capstick and N. Isgur: Phys. Rev. {\bf D34} (1986) 2809.

\bibitem{Glozmann} L. Glozmann {\em et al.}:  Phys. Rev. {\bf C57} (1998) 
 3406; Phys. Rev. {\bf D58} (1998) 094030.

\bibitem{CAR99} F. Cardarelli {\em et al.}: Few Body Syst. Suppl. {\bf 11} 
 (1999) 66.

\bibitem{JLAB} M.K. Jones {\em et al.}: Phys. Rev. Lett. {\bf 84} (2000) 
 1398.

\bibitem{KP91} For a review on the $LF$ form of the dynamics, see B.D. 
 Keister and W.N. Polyzou: Adv. Nucl.  Phys. {\bf 20} (1991) 225 and F. 
 Coester: Progr. Part. and Nucl. Phys. {\bf 29} (1992) 1.

\bibitem{CAR95} F. Cardarelli {\em et al.}: Phys. Lett. {\bf B357} (1995) 
 267; Few Body Syst. Suppl. {\bf 8} (1995) 345.

\bibitem{CK95} S. Capstick and B. Keister: Phys. Rev. {\bf D51} (1995)
 3598.

\bibitem{ZGRAPH} L.L. Frankfurt and M.I. Strikman: Nucl. Phys. {\bf B148} 
 (1979) 107. G.P. Lepage and S.J. Brodsky: Phys. Rev. {\bf D22} (1980) 2157. 
 M. Sawicki: Phys. Rev. {\bf D46} (1992) 474.

\bibitem{Platchkov} S. Platchkov {\em et al.}: Nucl. Phys. {\bf A510} (1990) 
 740.

\bibitem{Isgur99} For a recent review of the effects of the spin-spin force 
 among constituent quarks, see N. Isgur: Phys. Rev. {\bf D59} (1999) 034013 
 and references therein quoted.

\bibitem{Carbonell} J. Carbonell {\em et al.}: Phys. Rept. {\bf 300} (1998) 
 215 and references therein quoted.

\bibitem{CAR_rho} F. Cardarelli {\em et al.}: Phys. Lett. {\bf B349} 
 (1995) 393.

\bibitem{CAR_Delta} F. Cardarelli {\em et al.}: Phys. Lett. {\bf B371} 
 (1996) 7; Nucl. Phys. {\bf A623}(1997) 361.

\bibitem{Karmanov} V.A. Karmanov and J.-F. Mathiot: Nucl. Phys. {\bf A602} 
 (1996) 338.

\bibitem{Moshinsky} T.A. Brody and M. Moshinsky: {\em Tables of 
 transformation brakets}, Monografias del Instituto de Fisica, Mexico, 
 1960.

\end{thebibliography}
\end{document}